\documentclass[11pt]{article}
\usepackage{times}

\usepackage{amssymb}

\makeindex

\begin{document}
\def\alt{\lesssim}
\def\agt{\gtrsim}
\def\lsim{\lesssim}
\def\gsim{\gtrsim}
\def\Br{\nonumber\\ }
\def\ee{{\mathrm e}}
\def\dd{{\mathrm d}}
\def\e {\epsilon}
\def\eps {\varepsilon}
\def\D {{\cal D}}
\def\cS {{\cal S}}
\def\cF {{\cal F}}
\def\d {{\delta}}
\def\Sp{\textsf{Sp}\,}
\newcommand\U{\textsf{U}\,}
\newcommand\SO{\textsf{SO}\,}
\newcommand\SU{\textsf{SU}\,}
\def\S{\textsf{S}\,}
\def\Tr {\text{Tr}\,}
\def\tr {\text{tr}\,}
\def\s {{\sigma}}
\newcommand\ef{{\varepsilon_{\!_{\text{{ F}}}}}}
\newcommand\vf{{\upsilon_{\!_{\text{{ F}}}}}}
\def\l {{\lambda}}
\def\L {{\Lambda}}
\def\tfrac#1#2{{\mbox{$\frac#1#2$}}}
\def\SIGMA{\mbox{\ra{\large $\sigma$\ }1}}
\def\Eq#1{{Eq.~(\ref{#1})}}
\def\Ref#1{(\ref{#1})}
\def\Eqs#1{{Eqs.~(\ref{#1})}}
\def\EQS#1#2{{Eqs.~(\ref{#1}) and (\ref{#2})}}
\newcommand\EQ[2]{{Eqs.~(\ref{#1})--(\ref{#2})}}
\newcommand\E[1]{\!#1\!}
\newcommand\Fra[5]{\left #1 \frac {#3}{#4} \right #2^{\!#5}}
\def\lr#1#2#3{\left#1{#3}\right#2}
\newcommand\Int{\int\!\!}
\newcommand\spr[2]{{\mathbf {#1}} \cdot {\mathbf {#2}}}
\def\av#1{\left <{#1}\right >}
\def\ra#1#2{\raise #2pt\hbox{#1}}
\def\b#1{{\mathbf {#1}}}
\def\bra#1{\left\langle {#1}\right|}
\def\Lr#1#2#3#4{\left#1{#3}\right#2^{\!#4}}
\def\ket#1{\left| {#1}\right\rangle}
\def\tel{\tau_{\text{el}}}
\def\x{\hat{\xi}}
\newcommand\lF{{\lambda_{\text F}}}
\def\row#1#2{#1_1,\ldots,#1_{#2}}
\def\onlinecite#1{ \cite{#1}}
\def\text#1{{\mathrm {#1}}}
\def\bbox#1{{\mathbf {#1}}}
\def\NL{NL$\sigma$M}
\def\SM{$\sigma$ model}
\def\openone{\leavevmode\hbox{\small1\kern-3.3pt\normalsize1}}%

\title{Nonlinear Sigma Model for Disordered Media: Replica Trick for
Non-Perturbative Results and Interactions}

\author{ Igor V. {Lerner}\\\small School of Physics and
Astronomy, The University of Birmingham, Birmingham B15 2TT, UK}

\date{}
\maketitle

\abstract{In these lectures, given at the NATO ASI at Windsor
(2001),  applications of the replicas nonlinear sigma model to
disordered systems are reviewed. A particular attention is given
to two sets of issues. First, obtaining non-perturbative results
in the replica limit is discussed, using as examples (i) an
oscillatory behaviour of the two-level correlation function and
(ii) long-tail asymptotes of different mesoscopic distributions.
Second, a new variant of the sigma model for interacting electrons
in disordered normal and superconducting systems is presented,
with demonstrating how to reduce it, under certain controlled
approximations, to known ``phase-only'' actions, including that of
the ``dirty bosons'' model.  }

\section{Introduction}

Starting from the seminal papers of Wegner \cite{Weg:79} and
Efetov \cite{Ef:82}, a field-theoretical description based on the
nonlinear $\sigma$ model (NL$\sigma$M) has become one of the main
analytical approaches to various problems in disordered electronic
systems. The ensemble averaging over all configurations of
disorder is performed either using bosonic \cite{Weg:79,MK+S} or
fermionic \cite{EfLKh} $n$-replicated fields and taking the
$n\!\to\!0$ limit in the results, or using supersymmetric fields
\cite{Ef:82,VWZ}. The main advantage of this approach lies in
formulating the theory in terms of low lying excitations
(diffusion modes) which greatly simplifies perturbative and
renormalization group calculations, and also allows a
non-perturbative treatment.

The first application of this approach was a derivation
[1--4]
of the renormal\-ization-group (RG)
equations of the scaling theory \cite{AALR} of Anderson
localization. For such a perturbative approach, generalized later
for mesoscopic systems \cite{AKL:86}, both the replica and the
supersymmetric methods are equally justified: they  ensure the
cancellation of unphysical vacuum loops in a diagrammatic
expansion.

However, it soon became conventional wisdom that there exist two
sets of problems, for each only one of these methods being
applicable. The first set of problems  can be schematically
specified with the following TOE\footnote{\parbox[t]{14.2cm}{TOE stands for ``Theory
of Everything''. In order to be really a TOE model in a
condensed-matter context, the Hamiltonian \Ref{TOE} should also
include spin terms, as later in this paper in relation to
superconductivity. Naturally, either with or without spin terms,
there is no hope for a rigorous approach to, let alone the exact
solution of, this model for an arbitrary disorder/interaction
strength.}} model:
\begin{equation}
\label{TOE}
  H=\sum_i\frac{\hat p_i^2}{2m}+\sum_i V_i +\tfrac12
  \sum_{ij}V_{ij}\,.
\end{equation}
Here the second term represents schematically both lattice and
disorder potential, while the last represents all possible
two-particle interactions (including the BCS one where relevant).
To include the interactions, at least at the perturbative and RG
level,
 the original fermionic replica  \cite{EfLKh} has been
generalized by Finkelstein \cite{Fin}. The interest in this
approach has been greatly enhanced by the recent discovery
\cite{Krav:95} of an apparent metal-insulator transition in 2D
disordered systems in zero magnetic field. Naturally, the TOE
model and its $\sigma $ model implementation  covers a much wider
variety of systems and phenomena than in the scope of the original
NL$\sigma$M [1--4]
describing only non-interacting disordered systems.

After a few earlier attempts %
[9--11], it has been recently demonstrated \cite{AnKam:99,Lud:98}
 that the Keldysh technique \cite{Keld} provides a viable alternative to the
replica approach for interacting systems. However, the latter
still remains one of the best available tools for consideration of
interacting electrons in disordered systems. In the very least, it
is clear that there is no simple way of applying the supersymmetry
method to a many-particle {\it fermionic} system, so that the
choice is between using Keldysh technique or replica trick.

On the other hand, there exist another set of problems for
disordered electron systems where meaningful results are
essentially non-perturbative (or at least look like that -- see
section \ref{HighGradients} below). In the absence of
interactions, many of them have been solved  the help of the
supersymmetry method (see, e.g., Refs.\ [15--20]
out of many), while
the viability of the replica approach was seriously questioned.

 The first, and arguably most famous, of
such problems was solved by Efetov \cite{Ef:82b} who used the
supersymmetric NL$\sigma$M  to derive the two-level correlation
function (TLCF) in the universal ergodic regime for electrons in
disordered metallic grains. The results proved to be identical to
those for eigenvalue correlations \cite{RMT} in random matrix
theory (RMT), as had much earlier been conjectured by Gor'kov and
Eliashberg \cite{GE}. This was the first microscopic derivation of
essentially a non-perturbative result for noninteracting electrons
in disordered media which has opened the way to numerous new
results (again, see Refs.\ [15--20]
out of many -- this
paper is not a proper venue to give a comprehensive list, let
alone even a short review, of them).

In many ways,  this first non-trivial ``super\-sym\-metric''
result \cite{Ef:82b}
 seemed to be an excellent illustration of why the replica trick
could only be used within a perturbative approach. For the
(easiest) case of the unitary symmetry (one of Dyson's symmetry
classes corresponding to the absence of the time-reversal
symmetry, e.g.\ due to an external magnetic field)
 the irreducible TLCF is given by
\begin{equation}
\label{R2U} R_2(\omega ) = -\frac{\sin^2\!\omega }{\omega ^2}\,,
\end{equation}
where $\omega $ is the distance between two levels in units of $\d
/\pi$ and $\d $ is the mean level spacing. This result is valid in
the ergodic regime, i.e.\ for $\omega \ll g$ where $g\gg1$ is the
dimensionless conductance. For $\omega \gg 1$, the TLCF averaged
over fast oscillations could be readily obtained from the standard
diagrammatic techniques \cite{AS} with $1/\omega $ being the
perturbation parameter (or from the perturbation theory in the
framework of either the supersymmetric or replica NL$\sigma$M).
However, the non-perturbative factor $\sin^2\! \omega $ cannot be
restored from the perturbation series. Since the replica trick is
well justified only within the perturbative approach, it might
seem rather hopeless to obtain the result (\ref{R2U}) within the
replica approach. And indeed, quite involved calculations by
Verbaarschot and Zirnbauer \cite{VZ:85} have shown that a direct
application of the replica trick (using either the bosonic or
fermionic NL$\sigma$M) has apparently  not reproduced
 the TLCF given by Eq.~(\ref{R2U}).

In section \ref{replicas}, following Refs.\ [25--27],
I will show   that
introducing a proper replica-symmetry breaking within the
fermionic NL$\sigma$M leads to the non-per\-tur\-bative result
(\ref{R2U}), albeit only in the asymptotic region $\omega \gg1$.
 This raises hope that one
might eventually apply the replica approach for obtaining {\it
non-perturbative results} for interacting electrons in disordered
systems (see Ref.\ \cite{Kam:00} for the first steps in this
direction).

The replica-symmetry breaking is not the only way to get (at least
seemingly) a non-perturbative result from the replica NL$\sigma$M.
In section \ref{HighGradients} I will briefly summarise how quite
non-perturbative calculations [18--20]
(that could only be performed within the supersymmetric method)
have reproduced {\em part} of the results for long-range
asymptotics of different mesoscopic distribution  obtained much
earlier [29--32]
by using a perturbative RG
approach valid within any variant of the NL$\sigma$M.

Then, I will describe in section \ref{interactions} how to include
interactions into a modern variant of the replica \NL\  and will
use this model in section \ref{SIT} to describe different
approaches to superconductor-insulator transitions in thin
disordered films.

\section{Replica-symmetry breaking: a way to non-perturbative
results?}\label{replicas}

In this section we consider non-interacting electrons in
disordered media -- the problem described in a standard way by the
first two terms in Hamiltonian \Ref{TOE} (this is equivalent to
the Anderson model of disorder). The aim is to reproduce
non-perturbative oscillatory contribution to the TLCF of \Eq{R2U}
using the replica \NL.

Before describing the derivation, it is worth specifying what is
meant here by the replica-symmetry breaking. It is the standard
fermionic NL$\sigma$M which does contain correct  non-perturbative
oscillatory contributions to the TLCF. The replica symmetry is not
broken at the level of choosing a correct saddle-point
approximation but rather fully exploited by involving a set of
additional (in comparison to standard considerations \cite{EfLKh})
saddle-point submanifolds within the usual NL$\sigma$M. The
symmetry is broken in a rather natural way when accounting for the
contribution of these submanifolds to the TLCF. For the unitary
case, such a procedure gives the exact results for all $\omega$,
although a real reason for this \cite{Zirn:99} may be incidental.
There is no doubt, however, that it gives the asymptotically
($\omega\gg1$)  correct results for all the three Dyson's symmetry
classes.

Note an important analogy: in the supersymmetric NL$\sigma$M
Andreev and Altshuler \cite{A2} have reproduced the large-$\omega
$ limit of Eq.\ (\ref{R2U}) by the saddle-point calculation of the
supersymmetric integral that involved an additional saddle-point
which ``breaks'' the supersymmetry in a similar way. The  same
integral has been exactly calculated by Efetov \cite{Ef:82b}
without any breaking of the supersymmetry.

This analogy will be extended by showing (with the help of a trick
resembling the replica-symmetry breaking in the meaning explained
above) that the $n=0$ replica limit of the {\em exact} integral
representation of the TLCF obtained in Ref.\ \cite{VZ:85}  for the
fermionic replica NL$\sigma$M  leads to the same asymptotically
correct oscillatory behavior of \Eq{R2U}.

\subsection{Formulation of the problem in terms of \NL}

Consider the two-level correlation function (TLCF) defined by
\begin{equation}
\label{TLCF}
R_2(\omega ) = \frac1{\nu^2}\left<\nu(\eps + \omega ) \nu( \eps
)\right>-1\,.
\end{equation}
Here $\left<\ldots\right>$ stands for the ensemble averaging,
i.e.\ averaging over all the realizations of the impurity
potential $V(\b r) $ in Hamiltonian \Ref{TOE},  $\nu(\eps )$ is
the electronic density of states per unit volume defined in terms
of the spectrum $\{\eps _\alpha\}$ for a given sample as $\nu(\eps
) = L^{-d} \sum_\alpha\delta(\eps - \eps _\alpha)$,
$\quad\nu\equiv \left<\nu(\eps ) \right> =1/L^d \d $, and $L$, $d$
are the sample size and dimensionality. All energies are measured
in units of $\d/\pi$, in which the TLCF has the form of
Eq.~(\ref{R2U}). In these units  $R_2$ is expressed via the
product of the retarded and advanced Green's functions \cite{AGD}
as follows:
\begin{equation}
\label{S2} R_2(\omega)=\frac{1}{2}\left[\Re\text{e}\,
S_2(\omega)-1\right]\,, \quad  S_2(\omega)\equiv \left< G^{\,r}\!
\lr(){\eps +\frac{\omega}{2}} G^{\,  a}\!\lr(){\eps
-\frac{\omega}{2}} \right>\,.
\end{equation}
The function $S_2$ can be expressed in the standard way
\cite{Ef:82b,SLA} in terms of a generating functional $Z_n$
\begin{eqnarray}
\label{S2SM} S_2(\omega)&=&-\lim_{n\to
0}\frac{1}{n^2}\frac{\partial^2}{\partial\omega
^2}\lr<>{Z_n(\omega)},\quad Z_n=\Int\D\bar\psi\D\psi\, \ee^{i\cS}
\\[6pt]
\cS[\bar\psi,\psi]&=&\Int\dd \b r\,\bar\psi (\b r)\!
\left[\hat{\xi}-\lr(){\frac\omega2+i\eta}\Lambda+V(\b
r)\right]\!\psi (\b r)\,,\quad \eta\to+0\,.\label{Zn}
\end{eqnarray}
Here $\hat \xi=\hat p^2/2m-\ef$, $\bar\psi_\sigma$ and
$\psi_\sigma$ are conjugate $2 n\,$--\,component fermionic
(Grassmannian) fields (corresponding to $n$ replica components and
$2$ retarded-advanced components -- further $2$ spin components
are redundant in this section but will be included in
section~\ref{interactions}),
$\Lambda\equiv\mbox{diag}(1,-1)_{ra}\otimes\openone_n$.
 \footnote{\parbox[t]{14.2cm}{The same notation for $\L$ will be used in all extended
 spaces introduced later, assuming that it is a dyadic product of Pauli's $\s_z$
 in the RA space and of the unit matrices in
 all other sectors of the extended space.}} \
The replica trick, $n\E\to0$,  allows one to write the above
expression for the product of Green's functions in terms of
derivatives of $Z$ (rather than of $\ln Z$) which makes the
ensemble averaging straightforward. Performing it in the
assumption that $V(\b r) $ is a Gaussian $\delta$-correlated
random  potential with the dispersion $(2\pi \nu
\tau_{\text{el}})^{-1}$, where $\tau_{\text{el}}$ is the mean free
time of elastic scattering from impurities, one obtains the
following quartic term standing in the action $\cS$ for $\bar \psi
V\psi$:
 \begin{equation}\label{Sdis}
\cS_{dis}=-\frac{1}{4\pi\nu\tau_{\text{el}}}\int\dd{\bf r}\,\lr[]{
\spr{\bar\psi (\b r)}{\psi (\b r)}} \lr[]{\spr{\bar\psi (\b
r)}{\psi (\b r)}}
\end{equation}
The next step is the standard Hubbard--Stratonovich decoupling
that allows one to single out all ``slow'' modes of electron
motion. In the absence of any symmetry breaking (due to an
external magnetic field, or magnetic impurities, or spin-orbit
coupling), there are two modes, equally contributing to quantum
interference effects: the diffuson mode, corresponding to small
transferred momenta, and the cooperon mode, corresponding to small
sum momenta. It is instructive to perform the decoupling directly
in the $\b r $ space, thus emphasizing that the diffuson mode
corresponds to the  normal pairing  while the cooperon -- to the
anomalous pairing:
\begin{equation}
\begin{array}{l}
\displaystyle \exp\{-S_{dis}\} = \int\frac{\D d}{{\cal N}_d}
\exp\lr\{\}{\!-\!\Int\dd \b r \left[
\frac{\pi\nu}{4\tau_{\text{el}}}\tr\, d_\omega^2(\b r) -
\frac{i}{2\tau_{\text{el}}}
 \bar\psi d(\b r) \psi\right]}+ \\ [4mm]
\displaystyle \int\frac{\D c}{{\cal N}_c}
\exp\lr\{\}{\!-\!\Int\dd\b r\left[
\frac{\pi\nu}{4\tau_{\text{el}}}\tr\, c_\omega^+(\b r)c(\b r) -
 \frac{i}{4\tau_{\text{el}}} \left[
\bar\psi  c_\omega(\b r) \bar\psi - \psi c_\omega^+(\b r) \psi
\right]\right]}
\end{array}\label{dis}
\end{equation}
The diffuson field $d_\omega (\b r) $ is a matrix field of the
same symmetry as $\psi\otimes \bar \psi $ while the cooperon field
is of the same symmetry as $\bar\psi\otimes \bar \psi $. These two
fields can be combined into the  single $4N\times 4N$ matrix field
$\sigma(\b r)$ of the form

\begin{equation}
\label{dc}
 \displaystyle \sigma = \left(
\begin{array}{cc}
d & c \\
c^+ & d^T
\end{array}
\right), \quad \mbox{where} \quad d^+ = d, \,\,\, c^T = -c.
\end{equation}
The extra $2\times2$ ``diffuson-cooperon'' space, explicit in
\Eq{dc}, is usually referred to as the ``charge conjugation''
space (sometimes, as the time-reversal space): apart from being
Hermitian, the matrix field $\sigma$ has the charge conjugation
symmetry,
\begin{equation}\label{CG}
\sigma = \bar{\s}\equiv{\cal C}\s^* {\cal
C}^{-1}\,,\qquad\bar{\Psi} = \left({\cal C}\Psi\right)^T
\end{equation}
 where ${\bar \Psi} =
\frac{1}{\sqrt{2}} \left(\psi^+,-\psi^T\right), \quad \Psi^T =
\frac{1}{\sqrt{2}} \left(\psi^T,\psi^+\right)$, so that the second
equation in \Ref{CG} defines the charge-conjugation matrix $\cal
C$.  The disorder functional \Ref{dis} can be written in such an
extended space as follows:
\begin{equation}
\displaystyle \exp\{-S_{dis}\} = \int\frac{\D\sigma}{{\cal
N}_{\sigma}} \exp\left\{\Int\dd\b r\lr[]{
-\frac{\pi\nu}{8\tau_{\text{el}}}\tr\sigma^2+\frac{i}{2\tau_{\text{el}}}{\bar
\Psi}\sigma\Psi }\right\}\label{Dis}
\end{equation}
The functional integration measure in \EQS{dis}{Dis} is irrelevant
in the replica limit. After performing the Gaussian functional
integration over the fermionic fields in \Eq{S2SM}, one reduces
the generating functional $Z_n$ to $Z_n=\av{\ee^{-{\cF(\s)}}}_\s$
with $\av{\ldots}_\s$ standing for the functional integration over
all the independent components of the field $\s$, and the ``free
energy'' is
\begin{eqnarray}
\label{trln} {\cF}[\s]=\frac{1}{L^d}\int \text{d}^dr
\left\{\frac{\pi\nu}{8\tau_{\text{el}}}\tr\sigma^2-\frac12\tr
\ln\lr[] {\lr(){\frac\omega 2+i\delta}\L -\hat \xi +\frac i
{2\tau_{\text{el}}}\s } \right\}\,.
\end{eqnarray}
A saddle-point condition for this functional, found by varying the
field $\s$ is
\begin{eqnarray}
\label{SP}-i\pi \nu \s =\bra r\hat G\ket r\,,\qquad\hat
G\equiv\Lr[] {\lr(){\frac\omega 2+i\delta}\L -\hat \xi +\frac i
{2\tau_{\text{el}}}\s }{-1}\,.
\end{eqnarray}
In the class of spatially-homogeneous fields $\s$, this condition
is obviously satisfied by the field $\s=\L$ as can be easily seen
by transforming \Eq{SP} into the reciprocal momentum space and
performing integration over all momenta in the r.h.s.\ of this
equation in the pole approximation, justified provided that
\begin{equation}\label{eft}
 \ef\tel\gg1
\end{equation}
What is usually not stressed is that the saddle-point condition
\Ref{SP} is satisfied in the same approximation not only by $\L$
but by any spatially-homogeneous diagonal matrix commuting with
$\L$ with eigenvalues equal to $\pm1$. For most applications (but
not for what follows) the existence of this wide class of the
saddle-point solutions at $\omega\ne0$ is irrelevant. The reason
is that for $\omega=0$ \Eq{SP} is satisfied by a much wider class
of matrices, $\s=Q$, where $Q$ satisfies the following
constraints
\footnote{\parbox[t]{14.2cm}{It is worth stressing that the
condition $\tr Q=0$ arises from choosing an equal number of
replicas in both retarded and advanced Green's functions $G^r$ and
$G^a$.  With this choice, non-zero values of  $\tr Q$ would
correspond to ``massive'' modes that will not be contributing to
the results.}}
\begin{eqnarray}
Q^2=\openone _{2n}\,,\qquad\quad \tr Q=0\,. \label{Q2=1}
\end{eqnarray}
These constraints  are resolved by representing $Q$ as follows:
\begin{equation}
\label{UQU} Q=U^\dag \L U=T^\dag \L T\,,\qquad T
=\exp\left(\begin{array}{cc}0_{n}&t\\-t^\dag &0_n
\end{array}\right)\,.
\end{equation}
Since $Q$ obeys the charge conjugation condition \Ref{CG},  $U$ is
a symplectic matrix (a unitary matrix whose elements are real
quarternions \cite{EfLKh}), i.e.\  $U\in \Sp(2n)$. Then $T$ is
obtained by factorizing matrices $U$ with respect to redundant
matrices $R\in\Sp (n)\times \Sp (n)$ that commute with $\L $,
i.e.\ $U=RT$, which reduces $T$ to the form given in \Eq{UQU} with
$t$ being an arbitrary $n\!\times\! n$ real-quarternionic matrix.
This means \cite{DFN} that $Q$ belongs to the {\it compact}
Grassmannian manifold (coset space), $\Sp (2n)/\Sp (n)\times \Sp
(n)$. This class of symmetry corresponds to Dyson's orthogonal
class in RMT. I will discuss other symmetry classes slightly
later.

The final step of the \NL\ derivation is the expansion of $\tr
\ln$ in \Eq{trln} in small $\omega$ and $\nabla Q$ around a
spatially-homogeneous zero-frequency saddle point of
\EQS{Q2=1}{UQU}. To this end, one performs the similarity
transformation in the $\tr \ln$-term representing it as $$\tr \ln
\lr\{\}{G_0^{-1}-U[\hat \xi,U^\dag]+\frac12\omega U\L U^\dag
}\,,$$ where $G_0$ is the Green's  function of \Eq{SP} at
$\omega=0$ and $\s =\L$. After this, expanding the $\tr \ln$ to
the first power in $\omega$ and the second power in the gradient
operator $\hat\xi\equiv\hat {\b
p}^2/2m-\ef\approx\vf\spr{n}\nabla$, performing the pole
integration over $\xi$, justified in the already used
approximation $\ef\tel\gg1$, and neglecting in the same
approximation $\tr \s^2$ term in \Eq{trln}, one finally arrives at
the non-linear sigma model functional
\begin{eqnarray}
\label{nlsm} {\cal F}[Q;\omega]=\frac{1}{L^d}\int \text{d}^dr
\hbox{Tr}\left[\frac{1}8D(\nabla Q)^2-\frac{i\omega
\alpha}4\Lambda Q\right]\,,
\end{eqnarray}
where $D\E=\vf^2\tel/d$ is the diffusion coefficient; at $d\E=2$,
the dimensionless conductance $g \E= 2\pi^2\nu D$.

 If  time reversal invariance is
broken by a magnetic field or magnetic impurities, or
spin-rotation symmetry is broken by the spin-orbit interaction,
the above functional remains the same but the matrix $U$ in
\Eq{UQU} belongs, respectively, to the unitary group $\U (n) $
(Dyson's unitary class) or to the {\it orthogonal} group, $\SO
(n)$ (Dyson's symplectic class). Here $\alpha\E=1$ for the
orthogonal class and $\alpha\E=2$ for the unitary  and symplectic
classes. This factor arises because unitary and symplectic classes
have been obtained from orthogonal \cite{EfLKh} by the suppression
of massive modes corresponding to the time-reversal or
spin-rotational symmetry breaking and a subsequent reduction of
the $Q$ matrix rank.
 The coefficient $\alpha$ also absorbs an extra factor
in the symplectic case due to the redefinition of the mean level
spacing $\d $ in the chosen units.

For simplicity, I consider in the remaining part of this section
only the unitary class, referring to the original publication
\cite{KamMez:99b,YL:99a} for the other two classes. I will also
limit considerations to the ergodic regime corresponding to the
level separations much smaller than the Thouless energy,
$E_{\text{\sc T}}\sim D/L^2$ (which in the chosen units coincides,
up to a numerical factor, with the dimensionless conductance $g$).
In this regime the gradient term in Eq.~(\ref{nlsm}) may be
neglected, and the NL$\sigma$M functional reduces to the
zero-dimensional limit: \cite{Ef:82b}
\begin{equation}
\label{0d} {\cal F}[Q;\omega]= -\frac{i\omega }2\Tr\left[ \Lambda
Q\right]\,,
\end{equation}
with $Q$ becoming a spatially homogeneous matrix.

\subsection{Calculating the asymptotics of the TLCF}

\label{sec:unitary}

Limiting all further considerations to the ergodic regime only,
$Z(\omega )$ given by Eqs.\ (\ref{S2SM}) and (\ref{0d}) can be
represented as
\begin{equation}
\label{UZ} Z_n(\omega)=\int\!\D Q\exp\left[-i\frac{\omega}{2}\Tr\L
Q\right]\,,
\end{equation}
where the measure is defined by
\begin{equation}
\label{mes}
\D Q = \prod_{i,j=1}^n \dd \Omega _{ij}^{ra}\, \dd \Omega _{ij}^{ra\,*}\,,
\qquad d\Omega  \equiv dT\cdot T^{-1}\,.
\end{equation}
Here $T$ is the matrix parameterizing $Q$, Eq.~(\ref{UQU}), and
$r$ and $a$ refer to the replica indices which originate from
$G^r$ and $G^a$, respectively. In the large-$\omega $ limit this
integral is mainly contributed by the extrema of the functional
which obey the standard condition $ [\L, Q]=0$. This condition is
satisfied by any matrix of the form $Q= {\text{diag}}(Q^r, Q^a)$,
where $Q^{r}$ and $Q^{a}$
 are the $n\times n$ Hermitian
matrices whose eigenvalues are $\pm1$ and $\Tr (Q^r+ Q^a)=0$.
This defines a highly degenerate saddle-point manifold which consists of
$C_{2n}^n$ submanifolds specified by a particular distribution
of $n$ eigenvalues `+1' and $n$ eigenvalues `-1' between $Q^{r}$ and $Q^{a}$.
These submanifolds can be divided into
$n+1$ classes of equivalence, $Q_p=\text{diag}(Q^r_p, Q^a_p)$, labeled by
$\Tr Q^r_p=-\Tr Q^a_p=n-2p$, with $p=0,1,\ldots n$. The $p$-th class
has weight $(C_{n}^p)^2$, with $C_{n}^p\equiv {n \choose p}$.

The matrix $ Q^r_p$ with $(Q^r_p)^2=\openone_n$ and $\Tr Q^r_p=n\!-\!2p$
can be parameterized by analogy with Eq.~(\ref{UQU}) as
\begin{equation}
\label{Qrp}
 Q^r_p=( T^r_p)^\dag \l _p  T^r_p \,, \quad
\l _p\equiv \text{diag}( {\openone } _{n-p}, -\openone _p)\,,
\quad
  T^r_p =\exp\left(\begin{array}{cc}0_{n-p}&t^r\\-(t^r)^\dag &0_p
\end{array}\right)
\end{equation}
where $t^r$ is an arbitrary $p\times (n\!-\!p)$ matrix.
This defines
the coset space $\textsf{G}_p=\U(n)/\U(n\!-\!p)\!\times\! \U(p)$, i.e.\
Therefore,
$Q_p = \text{diag}(Q^r_p, Q^a_p)$ belongs to the manifold $\textsf{G}_p\times
\textsf{G}_p$ and can be parameterized
as
\begin{equation}
\label{Qp}
Q_p=T_p^\dag \L _p T_p\,,\qquad T_p=\text{diag}(T^r_p, T^a_p)
\,,\qquad
\L _p = \text{diag}(\l _p, -\l_p)
\end{equation}
The integer $p$ specifies the replica-symmetry breaking, as it
describes the number of the $-1$ eigenvalues in each $Q^r$ block
(equal to the number of the $+1$ eigenvalues in each $Q^a$ block):
in the symmetry-unbroken case, $p=0$, and hence retarded and
advanced blocks, $Q^{r,a}$, contain only positive or negative
eigenvalues, respectively.

Now one needs to take into account contributions from `massive'
modes (with mass $\propto 1/\omega $, not to be confused with the
massive modes $\propto 1/\ef\tau $ neglected upon the derivation
of the NL$\sigma$M) in the vicinity of each manifold (\ref{Qp}).
In the large-$\omega $ limit these contributions may be considered
as independent and the partition function is then represented by
the sum of all of them:
\begin{equation}
\label{Z}
Z_n(\omega)=\sum_{p=0}^{n}\left(C_n^p\right)^2 \int\!\D Q
\exp\left[-i\frac{\omega}{2}\Tr \L Q_p \right],
\end{equation}
This expression is somewhat symbolic, as\ $Q_p\equiv U\L _p
U^\dag$, covers the entire symmetric manifold of the NL$\sigma$M,
Eq.~(\ref{UQU}), including all the massive modes: indeed, $\Tr U
\L _p U^\dag \L\to\Tr \L _p T^\dag \L T= \Tr \L _p Q$, where we
have substituted
 $U=RT$, as defined after Eq.~(\ref{UQU}).
This can be justified only as a perturbative (in $1/\omega $)
procedure: a possible overlapping of massive modes originated from
different manifolds is irrelevant in the large-$\omega$ limit, and
each of the integrals in the sum \Ref{Z} can be calculated
independently of the others.

Each term in the sum ~(\ref{Z}) contains both massive and massless modes.
Indeed, we have used above the factorization $U=RT$
with $R$ being block-diagonal matrices commuting with $\L $. The matrices
$T$ in $\Tr \L _p Q=\Tr \L T \L_p T^\dag$ still contain the subset
of matrices commuting with $\L _p$ that correspond to the massless modes.
Therefore, we need to parameterize $T$ in a way which enables us to factorize
out these massless modes and perform the integration over the massive ones.

The most suitable parameterization of $T$,
 analogous to that used in Ref.~\onlinecite{VZ:85}
for the bosonic NL$\sigma$M, can be obtained
by expanding the matrix exponent in Eq.~(\ref{UQU}). By introducing
matrix $B\equiv t ({t^\dag t})^{-1/2} {\sin\sqrt{t^\dag  t}}$,
we represent $T$ and thus $Q=T^\dag \L T$ as follows:
\def\SQ{{\cal R(BB^\dag)} }
\def\SQQ{{\cal R(B^\dag B)} }
\begin{equation}
\label{p1}
T =
\left(
\begin{array}{cc}
\displaystyle
\SQ&
 \displaystyle
 B  \\
\displaystyle
-B^\dag  &
\displaystyle
\SQQ
\end{array}
\right)\,, \; Q = \left(
\begin{array}{cc}
\openone_n -2BB^\dag  & B\SQQ\\ 
B^\dag \SQ 
& -(\openone_n -2B^\dag B)
\end{array}\right)\,,
\end{equation}
where ${\cal R}(X)\equiv \sqrt{{\mathrm I}_n-X}.$ The matrix $B$
in this parameterization is not unconstrained, though. The
$Q=Q^\dag$ condition is fulfilled only when the matrices $\SQ$ and
$\SQQ$ are Hermitian. This is so only when all the eigenvalues of
$BB^\dag $ and $B^\dag B$ do not exceed unity. Only under this
constraint does $Q$, parameterized as in Eq.\ref{p1}, still belong
to the coset space $\U(2n)/\U(n)\times \U(n)$. Nevertheless, this
parameterization is very convenient. First, the corresponding
Jacobian is equal to one \cite{YL:99a} so that the measure of
integration (\ref{mes}) can be written  as
\begin{equation}
\label{BB}
\D Q = \prod_{i,j}\dd B_{ij}\dd B^*_{ij}\equiv \D B\,.
\end{equation}
In addition, the representation of
all the exponents in the sum~(\ref{Z}) in terms of $B$ is also very simple,
$
\Tr \L _p Q=2(n-2p) - 2\Tr
 \l_p (BB^\dag  + B^\dag B)
$, so that we obtain:
\begin{eqnarray}
\label{int} Z_n(\omega)&=&\sum_{p=0}^{n}\left(C_n^p\right)^2\cdot
\ee ^{i\omega(2p-n)}\,
Z_n^p(\omega)\,,\\
Z_n^p(\omega)&=& \int \!\D B \exp\left[i\omega\Tr \l_p (BB^\dag  +
B^\dag B) \right]. \label{Znp}
\end{eqnarray}
The region of integration in (\ref{Znp}) is restricted by the
constraint described after Eq.~(\ref{p1}). Last, but not least,
the parameterization (\ref{p1}) allows one to separate out the
massless modes, which obey the condition $[T, \L _p]=0$, in each
integral (\ref{Znp}). Indeed, this condition is satisfied by all
matrices $T$ constructed from $B$ which anticommute with $\l _p$,
i.e.\ have the off-diagonal block structure.

This means that in the representation of $B$ in the block form
reflecting the structure of $\l_p={\mathrm {diag}}(\openone_{n-p},
-\openone_p)$,
\begin{equation}
B = \left(
\begin{array}{cc}
  B_1   &  b_1 \\
 b_2^\dag    &  B_2
\end{array}\right),
\label{Bb}
\end{equation}
the matrices $B_{1,2}$ represent the massive modes, and $b_{1,2}$ massless.
When the massive modes are suppressed ($B_1=0$ and
$B_2=0$), the $T$ matrices in Eq.~(\ref{p1}) constructed from
$p\times(n\!-\!p)$ matrices $b_{1,2}$
only, parameterize the same degenerate $p$-the manifold, $\textsf{G} _p
\!\times\! \textsf{G} _p$ described in Eq.~(\ref{Qp}), as one expects.

By substituting the representation (\ref{Bb}) into
Eq.~(\ref{Znp}), we reduce $Z_n^p$ to the product of integrals over
the massive and massless modes:
\begin{equation}
\label{comp} Z^p_n(\omega)= \int \D B_1\, \D B_2
\exp\left[-2i\omega\tr(B_1 B_1^\dag  - B_2 B_2^\dag )\right] \int
\D b_1 \D b_2 ,
\end{equation}
Here the region of
integration over $b_{1,2}$ depends on $B_{1,2}$ due to the constraint
on the eigenvalues of the matrices $B B^\dag $ and
$ B^\dag B$ in the representation (\ref{Bb}).
 Since the
integral over $B_{1,2}$ is contributed only by the region where
both $\tr B_1 B_1^\dag $ and $\tr B_2 B_2^\dag \alt1/\omega \ll1$,
in the leading in $1/\omega$ approximation we may put both
$B_{1,2}$ to $0$ in the constraint of the integration
 region over the massless modes $b_{1,2}$. In this approximation, as we
have noticed after Eq.~(\ref{Bb}), matrices $b_{1,2}$ parameterize the
$p$-th manifold (\ref{Qp}) so that
\begin{equation}
\label{bb} \int \D b\equiv\int \D b_1 \D b_2 =\int \D
Q_p=\Omega^2(\textsf{G}_p)\,,
\end{equation}
where the measure of integration over $\D Q_p$ is defined in terms of $T_p$
in the same way that $\D Q$ is defined in terms of $T$, Eq.~(\ref{mes}),
and $\Omega(\textsf{G}_p)$ is the volume
of the compact coset space $\textsf{G}_p$. This volume is expressed
via the well-known volumes of the unitary group, $\Omega(\U (n))$, as follows:

\begin{eqnarray}
\label{vu}
\Omega(\textsf{G}_p)&=&\frac{\Omega(\U(n))}{\Omega(\U(n-p))\Omega(\U(p))}\Br&=&
(2\pi)^{\frac{1}{2}[n^2-(n-p)^2-p^2]}\prod\limits_{j=1}^{p}
\frac{\Gamma(1+j)}{\Gamma(n + 2 - j)}.
\end{eqnarray}
In the same large-$\omega $ approximation,
 the variables $B_{1,2}$ parameterizing the massive modes
are unconstrained. Then the Gaussian integral over the $2[(n\!-\!p)^2 + p^2]$
independent massive modes yields
\begin{equation}
\label{GI} \tilde Z^p_n(\omega)\equiv \int\!\! \D B
\exp\left[2i\omega\tr(B_1 B_1^\dag  - B_2 B_2^\dag )\right]=
\left(\frac\pi{-i\omega } \right)^{\!\!(n-p)^2}
\!\!\left(\frac\pi{i\omega } \right)^{\!\!p^2}
\end{equation}
Combining Eqs.~(\ref{GI}) and (\ref{vu}) and omitting
 an irrelevant overall factor which goes to $1$
when $n\E\to0$, we arrive at the following expression which is
essentially the same as that derived in
Ref.~\onlinecite{KamMez:99b} via the Itzykson-Zuber integral:
\begin{equation}
\label{un}
Z_n(\omega)=\sum_{p=0}^{\infty}\left[F_n^p\right]^2\cdot \frac{\ee
^{i\omega(2p-n)}}{(2\omega)^{(n-p)^2+p^2}}, \qquad F_n^p\equiv
C_n^p \prod\limits_{j=1}^{p}\frac{\Gamma(1+j)} {\Gamma(n+2-j)}.
\end{equation}
Here the summation over $p$ has been extended to $\infty$ since
$F^p_n=0$ for all integer $n>p$. This allows one to take
the replica limit, $n\to0$, in each of the terms in Eq.~(\ref{un}).
Due to the fact that $F^p_n\propto n^p$ as $n\to 0$,
only the terms with $p=0$ and $p=1$
 in Eq.~(\ref{un}) contribute to $S_2$ in Eq.~(\ref{S2SM}).

Let us stress that the replica symmetry is broken only now, in the $n\to0$
limit.  Indeed,
for any integer $n\ne0$ contributions of the  terms with $p$ and $n-p$ are
complex conjugate to each other, but for  $n\to0$
 we no longer treat them on equal footing.
Thus, the result for $Z_n(\omega)$ below is no longer a real
function. Note, however, that in order to treat $S_2(\omega) $ for
all $\omega$ one should imply the $\omega\to\omega +i\delta$
substitution which results in $Z_n(\omega)$ being no longer a real
function for any $n$.

Omitting all the terms with $p\ge2$, one obtains
\begin{equation}
\displaystyle Z_n(\omega)=\frac{\ee ^{-i\omega n}}{\omega^{n^2}} +
n^2 \frac{\ee ^{i\omega(2-n)}}{4\omega^{(n-1)^2+1}}\,.
\end{equation}
Substituting this into Eqs.~(\ref{S2SM}) and (\ref{S2}) and
keeping the leading in $1/\omega $ terms only, one arrives at the
expression (\ref{R2U}) for the TLCF. Although this expression
coincides with the exact one (implying the above mentioned
substitution $\omega\to\omega +i\delta$), it has been actually
derived only in the large-$\omega $ limit, as is the case of the
`supersymmetry breaking' method  of Andreev and Altshuler
\cite{A2}.

\subsection{The large-{\large $\omega $} limit of the
Verbaarschot-Zirnbauer Integral} \label{sec:VZ}

In order to obtain an explicit multiple-integral representation of the
 `zero-mode' partition  function, Eq.~(\ref{UZ}),
one uses the following `polar' decomposition of $Q$:
$$
Q=\left(
\begin{array}{cc}
u_1^+ & \\
& u_2^+
\end{array}
\right) \left(
\begin{array}{cc}
\lambda & \sqrt{1-\lambda^2}e^{i\phi}\\
\sqrt{1-\lambda^2}e^{-i\phi}& -\lambda
\end{array}
\right) \left(
\begin{array}{cc}
u_1 & \\
& u_2
\end{array}
\right),
$$
where $-1\leq \lambda_i \leq 1$ and $u_{1,2}$ are unitary matrices. The
appropriate measure of integration is given by
$$
{\mathrm D}Q=\prod_{i<j}(\lambda_i-\lambda_j)^2 \prod_{i}
\dd\lambda_i\,\dd\phi_i\, \dd\mu (u_1)\, \dd\mu (u_2)\,.
$$
The action in Eq.~(\ref{UZ}) depends only on $\lambda$, and the
integrations over $\dd\mu (u)$ give the volumes of the appropriate
unitary group. Thus one obtains
\begin{equation}
\label{VZ} Z_n(\omega)=\Omega^2(\U(n))\int\limits_{-1}^{+1}
\d^2(\l)\, \prod_{i=1}^{n} \ee ^{-i\omega\l_i}\text{d}\l_i \,,
\qquad \d(\l) \equiv \prod_{i<j}(\l_j - \l_i)\,.
\end{equation}
This is equivalent (with accuracy up to factors going to $1$ in
the $n\to0$ limit) to the representation for $S_2$ given in
Eq.~(2.24) of the paper by  Verbaarschot and Zirnbauer,
 \cite{VZ:85} which has been used for the critique of the replica
trick. We will show that it leads, in the very least, to the exact
large-$\omega$ assymptotic behavior of $S_2(\omega)$.

The leading in $1/\omega$  contributions to this highly
oscillatory integral (which does not have stationary points inside
integration region) come from the end points. To single out  these
contributions, we must take some $\l$'s close to $+1$ and the rest
close to $-1$, which imitates replica symmetry breaking. Let us
choose $n\!-\!p$ of $\l $'s  close to $+1$ and $p$ of $\l $'s
close to $-1$. Then we can  split up the Vandermonde determinant
in the following way:
$$
\d^2(\l ) =\prod_{i,j}^n|\l_j - \l_i|\approx
2^{2p(n\!-\!p) } \Delta^2_+\,\Delta^2_-
$$
where
\begin{equation}
\Delta^2_+ = \prod_{i,j=1}^{n-p}|\l_i-\l_j|\, \quad
\Delta^2_- = \prod_{i,j=n-p+1}^{n}|\l_i-\l_j|.
\end{equation}
Reducing the integral (\ref{VZ}) to the sum of such contributions,
we represent it as
\begin{eqnarray}
Z_n(\omega)\approx\Omega^2(\U(n))\sum_{p=1}^{n}(C_n^p)\,2^{2p(n-p)}
\int\limits_{-\infty}^{+1} \Delta^2_+\, \prod_ {j=1}^{n-p} \ee
^{-i\omega\l_j}\,\dd \l_j\;\nonumber\\\times
\int\limits_{-1}^{+\infty}\Delta^2_- \!\! \! \prod_ {j=n-p+1}^n
\ee ^{-i\omega\l_j}\,\dd \l_j\,.
\end{eqnarray}
Since in each of the integrals all the variables are close to one of
the limits of integration, the second limit was extended to
infinity.
Now we make substitutions
 $\l_i = 1-x_i$ in the first integral,  and $\l_i = -1+x_i$ in the
second one, reducing the above sum to the form:
\begin{equation}
\label{sum}
Z_n(\omega)\approx\Omega^2(\U(n))\sum_{p=1}^{n}(C_n^p)\,2^{2p(n-p)}
\ee ^{i\omega (2p\!-\!n)} I_{n\!-\!p} I_p
\end{equation}
where $I_p$ are integrals of Selberg's type: \cite{RMT}

\begin{equation}
I_p=\int\limits_{0}^{\infty}\Delta^2(x)\prod_{j=1}^{p}\text{d}x_j
\text{e}^{-i\omega x_j}
\end{equation}
Substituting the known Selberg integrals and
 discarding an overall factor
which goes to unity in the replica limit we arrive at
\begin{equation}
\displaystyle Z_n(\omega)=\sum_{p=0}^{n}\left[F_n^p\right]^2\cdot
\frac{\ee ^{i\omega(2p-n)}}{2\omega^{(n-p)^2+p^2}}, \quad F_n^p=
C_n^p \prod\limits_{j=1}^{p}\frac{\Gamma(1+j)} {\Gamma(n+2-j)}.
\end{equation}
This expression is exactly the same as Eq.~\ref{un} obtained in
section~\ref{sec:unitary} with the help of the replica-symmetry breaking.
Therefore, the exact representation (\ref{VZ}) {\it does contain}
the true oscillatory asymptotic behavior of the TLCF.

The authors of Ref.~\onlinecite{VZ:85} have also drawn attention
to the fact that there is an apparent contradiction between the
$\omega =0$ limit for $S_2$ obtained from the replica trick and
the exact supersymmetric result. Indeed, if $\omega$ is put to $0$
in the expression for $S_2$ following from Eqs.~(\ref{VZ}) and
(\ref{S2SM}) the $n\to0$ limit is taken {\it after that}, one
obtains $S_2(\omega\!=\!0)=-1$. This cannot be correct as
$\Re\text{e}\, S_2(\omega\!\to\! 0)>0$ as follows from the
definition (\ref{S2}) and, moreover, it is known that $
S_2(\omega\!\to\! 0)\to\delta(\omega)$. What is interesting,
however, is that if one separates
 the  singular, $S_2^{\text{sing}}(\omega)$, and
regular, $S_2^{\text{reg}}(\omega)$, parts of the exact
$S_2(\omega)$, then $S_2^{\text{reg}}(\omega\!\to\! 0)=-1$.
Therefore, the replica method gives
$S_2^{\text{reg}}(\omega\!=\!0)$ correctly, and it is just
$S_2^{\text{sing}}$  which is missing. However, the fact that
$Z_n(\omega \!\to\!0)$ is finite for any integer $n$ does not
necessarily implies that it is also finite (as a function of
$\omega\to0$) in the replica limit. For example, if the expansion
of $S_2 (\omega )$ {\it before} taking the $n=0$ limit contained a
term proportional to $\omega^{n^2-1}$, it would be singular in the
replica limit. In other words, if a non-trivial dependence on the
order of limits $n\to0$ and $\omega\to0$ existed, in a spirit of
the replica trick the $n\to0$ limit should be taken first. At the
moment, though, this remains only a speculation. However, the fact
that the large-$\omega $ limit of this integral reproduces the
correct results (\ref{R2U}) makes it plausible that there are only
technical difficulties rather than one of principle
 in the application of the replica method.

\subsection{The small $\omega$ limit} \label{sec:con}

The purpose of this section was to demonstrate explicitly that
non-perturbative oscillatory contributions to the TLCF of
electrons in a random potential could be extracted from
 the standard NL$\sigma$M
formulated in fermionic replicas\cite{EfLKh} many years ago. To
this end, all one needs is to parameterize all the non-trivial
saddle-point manifolds corresponding to the broken
replica-symmetry and describing `massless modes' of the theory,
and expand the action in the vicinity of these manifolds to
include `massive modes'. The very similar approach has been used
in the supersymmetric NL$\sigma$M: the non-perturbative
oscillations have been extracted by the expansion around two
extremal points one of which breaks the supersymmetry  \cite{A2}.
Since the exact supersymmetric calculation of the TLCF was well
known \cite{Ef:82b}, it was clear that the supersymmetry breaking
\cite{A2} was just a convenient method of extracting the
large-$\omega $ limit (and going beyond  the universal `zero-mode'
approximation). It has also shown in section \ref{sec:VZ} that the
exact integral representation of $R_2(\omega )$ does contain the
correct behavior in the large-$\omega $ limit.

The small $\omega$ limit is, arguably, more interesting as it
governs a number of non-perturbative results obtained within the
supersymmetric  NL$\sigma$M. I will briefly summarise some of them
[18--20] in the following section, emphasizing that they have
reproduced some of the results obtained much earlier
[29--32] by using a perturbative RG approach valid within any
variant of the NL$\sigma$M.

\section{Tails of distribution functions}
\label{HighGradients}

All extensive physical characteristics of disordered systems
fluctuate from sample to sample. However, only in mid-eighties it
was understood  that the scale of such fluctuations is governed by
the quantum coherence length so that at $T\to0$ they are not
reduced with increasing the sample size. Immediately after the
discovery \cite{Al:85,L+S:85} 
of the universal conductance fluctuations (UCF), it was shown
\cite{AKL:86} that their distribution function is Gaussian in its
bulk, but has long lognormal tails. The existence of such tails
appears to be a common feature of distribution functions of many
observable quantities [29--32] 
such as global and local density of states, different relaxation
times, etc. With increasing the disorder, the part of these tails
in distribution functions was increasing so that the entire
distribution of any local quantity was becoming lognormal in the
critical regime in the vicinity of the metal-insulator transition
\cite{IVL:88}. The characteristic feature of the lognormal
distribution is that the logarithm of its $s^{\text{th}}$ moment
is proportional to $s(s\!-\!1)\ln L$ where $L $ is the coherence
length (at zero $T$, this is just a sample size). Therefore, all
the moments (or in the limit of weak disorder, where only the
tails are lognormal, all the high moments) scale with different
exponents, i.e.\ system shows multifractal behaviour.

The results for the tails of the distribution functions that had
been originally obtained[29--32]
by the RG treatment of an extended (in a way described below) \NL,
have later been reproduced [18--20]
in a much more elegant way in the framework of the standard
supersymmetric \SM. I do not intend to describe this approach here
but only want to stress that some of the steps seem to be
absolutely impossible within the replica treatment. The $Q$ field
in the supersymmetric model belongs to a ``supermanifold'' that
includes both compact and a noncompact sectors. The existence of
the noncompact sector is absolutely crucial for finding the
distribution tails. With properly imposed boundary conditions
required for finding the tails, $Q=\L$ does no longer represents
the saddle-point. A correct saddle-point equation reduces, e.g.,
to the Liouville equation \cite{EF1} for a single scalar parameter
that parameterises the (only relevant) noncompact sector of the
space of the field $Q$ (instead of, e.g., the parameterisation
\Ref{p1} in terms of the unconstrained matrix $B$ or any other
matrix parameterisation required in the replica \SM). A
non-perturbative solution to this equation provides for the
existence of lognormal tails of the local density of states
distribution \cite{Mirl:96} or, equivalently, for the
multifractality of the wave function \cite{EF1}. Had such a
solution been found before the RG solution of
Refs.~[29--32],
it would be taken as yet another `example' of the incapacity of
the replica method. However, although the replica method works
perfectly within the RG approach, there is still no clear
understanding of why the results obtained within the
non-perturbative approach outlined above are exactly the same as
those within the RG approach to the extended \NL\ (in the limit of
the weak disorder, to which the validity of the non-perturbative
supersymmetric approach is limited).

\subsection{Local density of states in open systems}
The two-level correlation function considered above is not well
defined for $\omega\lsim g$ for an open system, as level widths
become of the order of Thouless energy, $E_{\text{\sc
T}}=\hbar/\tau_{\text{erg}}$ which is much bigger than the level
mean spacing $\delta$ (their ratio is of order $g\gg1$ in the
metallic limit). Its direct analog in this regime is the variance
of the density of states (DoS) at the Fermi energy. It is useful
to consider also the entire distribution of DoS which can be found
in terms of its irreducible moments (cumulants). The relative
values of the DoS cumulants can be expressed via the effective
field-theoretical functional in a way similar to that in
\EQ{S2SM}{Zn}:
\begin{equation}\label{DosCumulants}
R_s\equiv\frac{\av{\av{\nu^s}}}{\av \nu^s}=\lim_{n\to
0}\frac{(i)^{2s}}{\lr(){2n^2}^{s}}\frac{\partial^{2s}}{\partial\omega
^{2s}}\av{\lr<>{Z_n(\omega)}}\,.
\end{equation}
The cumulants of local DoS are given by the same expression with
partial derivatives substituted by variational ones, $\omega$
being considered in this case as a spatial-dependent source field
in \Eq{Zn}. As the higher derivatives with respect to $\omega$ are
involved in the above expression, the Tr ln in \Eq{trln} should
now be expanded to the higher powers of $\omega$. This leads to
the following additional contribution \cite{AKL:91} to the \SM\
functional \Ref{nlsm}:
\begin{eqnarray}
\label{nlsm-add} {\cal F}_{\text{add}}[Q;\omega]=-\frac{i
\alpha}4\sum_{s=1}^\infty\Gamma_s\Tr\left[\omega\Lambda
Q\right]^{s+1}\,,\quad
\Gamma_{s0}=\Fra(){i\tel\delta}{2\pi}s\,\frac{(2s\E-1)!!}{(s\E-1)!}\,.
\end{eqnarray}
As the bare value, $\Gamma_{s0}$, of the additional ``charge'' is
proportional to the $s^\text{th}$ power of the small parameter
$\tel\delta\sim(\lF)^{d\E-1}\ell/L^d$, their direct contribution
to the DoS cumulants is negligible compared to that obtained by
the repeated differentiation \Ref{DosCumulants} of the $\omega$
term in the standard \SM.\footnote{\parbox[t]{14.2cm}{In
calculating the local DoS cumulants, this small parameter is
compensated by the appearance of powers of $\delta(\b r
=0)\to\pi\nu/\tel$ so that the contribution of the additional
terms to the variance and higher cumulants of the local DoS is
comparable \cite{IVL:88} to that of the standard \SM.}}

However, the charges $\Gamma_s$ sharply increase under RG
transformations \cite{AKL:91}. While ``normal'' contribution to
the DoS cumulants, i.e.\ those from the standard \SM, \Eq{nlsm},
scale with the system size as appropriate powers of the DoS
variance, the additional contribution from the vertices
\Ref{nlsm-add} show the multifractal scaling due to this RG
increase of $\Gamma_s$. For Dyson's orthogonal class, it reads
\begin{equation}\label{add}
R_s^{\text{add}}\propto
g_0^{1-s}\Gamma_{s0}\,\ee^{us(s\E-1)}\,,\quad u\equiv \ln \frac{
\sigma_0}\sigma\approx g_0^{-1}\ln\frac L \ell\,,
\end{equation}
where $\sigma$ is the conductivity at the scale of the system size
$L$ and $\sigma_0$ is its bare value, that is the conductivity at
the scale of the mean free path $\ell$. As the small parameter
$\Gamma_{s0}$ scale as a linear power of $s$, for large enough $s$
the $\ee^{us(s\E-1)}$ factor becomes dominant.

The additional contribution to the  $s^\text{th}$ cumulant  of the
local DoS differs from that in \Eq{add} by the absence of the
small parameter $\Gamma_{s0}$. Such a contribution becomes
dominant for $s\gsim u^{-1}$ (or for $s\gsim g_0$ in the weak
disorder limit). The approximate equality in \Eq{add} refers to
the weak disorder limit. In this limit, the result for the
multifractal cumulants \Ref{add} has been reproduced within the
supersymmetric approach \cite{EF1,Mirl:96} outlined above. Such a
behaviour of the high cumulants leads to the lognormal tails of
the distribution function. With increasing the disorder, $u$
becomes of order $1$ in  the vicinity of the metal-insulator
transition in $d>2$, or for the region of strong localization in
$d=2$ dimensions. In this region, the entire distribution becomes
lognormal. The physical meaning of this is that there exist rare
realisations of disorder which gives values of the local DoS (or,
equivalently, local values of the wave function amplitude
$|\Psi(\b r)|^2$) which are much higher than the average value.
The importance of this untypical realisations greatly increases
with increasing the disorder so that they correspond to
prelocalized states.

\subsection{Is there a paradox?}

The properties of tails of different distributions governed by the
existence of the prelocalized states are described in detail in
reviews \cite{AKL:91,MirlinReview}. The point of this presentation
is to underline certain peculiarities of the replica method. To
this end, let me first outline how the RG results which lead to
\Eq{add} have been obtained. The composite vertices \Ref{nlsm-add}
are not closed under the RG transformations of the \SM\ given by
\EQS{nlsm}{nlsm-add}. The index-permutation symmetry lacking in
\Eq{nlsm-add} is restored under RG transformations via generating
the following additional vertices in each $s^{\text{th}}$ order:
\begin{eqnarray}
\label{nlsm-add2} \omega^{s+1}\prod_{l=1}^\infty\left[ \Tr\left(
\Lambda Q\right)^l\right]^{s_l}\,,\quad \sum_{l=1}^\infty
s_l=s+1\,.
\end{eqnarray}
Thus in each $s^{\text{th}}$ order the RG equations become matrix
equations, the matrix rank being equal to the number of partitions
of the integer $s$ into the sum of positive integers. These
equations are exactly solvable \cite{IVL:88} for any integer $n$,
and keeping only the largest eigenvalue leads in the replica limit
$n\to0$ to the result \Ref{add}. As the RG approach is purely
perturbative,\footnote{\parbox[t]{14.2cm}{\Eq{add} has been
obtained as a result of the one-loop RG analysis. The higher-loop
contributions limit its validity to $s\alt g_0^{3/2}.$ Such a
limitation ensures that no moment will be proportional to a
positive power of the system volume, $L^d$, i.e.\ all the
multifractal dimensionalities remain positive. A similar
restriction in the supersymmetric method of Ref.~\cite{EF1}
follows directly from the inapplicability of the standard \SM\
description at ballistic scales.}} the applicability of the
replica method is beyond doubt. However, there exist some apparent
contradiction.

The point is that one can choose different number of replicas for
to represent the advanced and retarded sectors. In the simplest
unitary case the $Q$ matrix would belong to the coset space
$\U(n)/\U(n\!-\!m)\!\times\! \U(m)$ where $n$ is the total number
of replicas, and $m$ is the difference in their numbers in the
advanced and retarded sectors. In the replica limit, $n\E\to0$ and
$m\E\to0$, the result is naturally the same as before. However,
the largest eigenvalue of the RG equations happens to depend only
on $n$ but not on $m$. Then, instead of taking the replica limit,
one could fix $m=1$ and take an arbitrary integer $n$. The
resulting model is defined on the coset space $\SU(n)/\U(n\!-\!1)$
isomorphic to sphere, i.e.\ in this limit the model becomes
equivalent to the $n$-vector model. The composite vertices
\Ref{nlsm-add} reduce in this case simply to $\cos^s\varphi$ where
$\varphi$ is the angle between the $n$-field and the direction of
an external (magnetic) field. It is well known (see, e.g.,
Ref.~\cite{Pokrovsky}) that such operators are irrelevant in the
$n$-field theory, i.e.\ they only decrease under the RG
transformations, in contrast to those of \Eq{nlsm-add}. So here is
the contradiction as they should correspond, as described above,
to a particular case of the matrix operators. It is clear, though,
that a direct correspondence is anyway impossible, as the rank of
the $Q$-matrices in \Eq{nlsm-add} and thus the number of the
eigenvalues in the RG equations depends on $m$ and in the
symmetric case $m\E=0$ it is equal to the number of partitions of
the integer $s$ into the sum of positive integers, i.e.\  much
larger than in the case $m\E=1$ when it is just equal to $s$.

The resolution of the apparent paradox is in the fact that one
should look only at correlation functions (observables) rather
than at the RG dimensions of the composite vertices. In
particular, only vertices $[\Tr \L Q]^s$ out of all the variety in
\Eq{nlsm-add2} contribute to the cumulants \Ref{DosCumulants},
while all the rest serve to give correct eigenvalues. In
calculating such a contribution, each eigenvalue enters with a
coefficient of proportionality which is a polynomial of $n$ and
$m$. It appears that all the polynomials attached to ``wrong" (in
the case $m\E=1$) eigenvalues vanish at all integer values of $n$
thus restoring the correspondence between the matrix and the
standard treatments of the $n$-model. Simultaneously, the
$n\E=m\E=0$ replica limit gives the results \Ref{add} discussed
above. On the face of it, it looks similar to the ``replica
symmetry breaking" discussed in the previous section: there exists
a set of coefficients vanishing at all integer $n$ and giving a
nonzero contribution at $n\E=0$. However, all the results
discussed here have been obtained within the perturbative RG
approach, and thus no trick was involved -- the replica limit here
is just to cancel unwanted vacuum loops in the RG diagrams. All
these RG results can be equally easy obtained within the
supersymmetric \SM.

One question that remains, though, is the following.  The
multifractal cumulants \Ref{add} have been obtained here within
the {\em extended} \SM, \EQS{nlsm}{nlsm-add}, while later they
have been reproduced within the standard SUSY \SM. So, does one
really need the additional vertices \Ref{nlsm-add}?

\subsection{High gradient vertices: RG or not RG?}
The answer to this question is in getting -- intermediately --
even more vertices. Those in \Eq{nlsm-add} have been obtained in
expanding the Tr ln term in \Eq{trln} in higher powers of
$\omega$. But it could be also expanded in higher powers of
$\nabla Q$ which yields the following set of vertices
\cite{KLY:88}:
\begin{equation}\label{Hgrad}
\cF_s \bigl[\,Q\bigr]\equiv \,\gamma_s
\eps_{\alpha_{_1}\ldots\alpha_{_{2s}}}
 \int\Tr\Bigl( \prod_{i=1}^{2s}\partial_ {\alpha_i}
Q\Bigr)\,\dd^dr\,.
\end{equation}
The renormalization of these vertices results in the same increase
of the largest eigenvalue as in \Eq{add}, and actually, taken
together with the $\omega$-vertex in the standard \SM\ of
\Eq{nlsm}, they contribute to the cumulants, \Eq{add}, in the same
way (but for unknown -- in any method -- pre-exponential factors)
as the composite vertices \Ref{nlsm-add}. Therefore, the latter
are not even necessary to obtain the multifractality.

An important point now is that, although the higher-gradient
vertices \Ref{Hgrad} have been obtained from the higher-order
expansion of the Tr ln term in \Eq{trln}, they are also generated
in performing the RG transformation of the standard \NL. Usually
they are omitted as being na\"\i{vely} irrelevant. However, they
become relevant at the one-loop level and actually govern
long-tail asymptotics of different distributions. Note that in
contrast to the composite vertices \Ref{nlsm-add}, the
high-gradient operators remain relevant (in the same sense) also
for the $n$ vector model \cite{Weg:90a}.

Now the possible relation between the RG approach outlined here
and the supersymmetric approach of Refs.\ \cite{EF1,Mirl:96} is
the following. The boundary conditions imposed to find correctly
the tails of the distributions in the latter approach makes the
saddle-point solution spatially inhomogeneous, while in the former
approach the relevance of the high gradients may indicate that
such a spatial inhomogeneity develops spontaneously when starting
from the standard homogeneous saddle-point.

Finally, there is a possibility that the role of the high-gradient
operators \Ref{Hgrad} is considerably more crucial. In the
perturbative RG approach, they have only contributed to the higher
moments of conductance, DoS, etc, since the structure of the RG
equations turns out to be triangular, i.e.\ the higher order
gradient terms do not contribute to the renormalization of the
lower order terms.  However, in contrast to the composite
operators \Ref{nlsm-add}, the  operators \Ref{Hgrad} do not break
any symmetry of the \NL. Therefore, one cannot exclude a
possibility of a non-perturbative contribution from these
operators to, say, the average conductance. Had such a
contribution existed, it would totally ruin the one-parameter RG
for the average conductance, and this the one-parameter scaling
description of the Anderson localization. At the moment, such a
possibility remains purely speculative.

\section{Coulomb and pairing interactions in the sigma model}
  \label{interactions}

  The main point of the previous sections is that the replica \SM\
works perfectly for perturbative problems and -- in certain cases
-- does reasonably well even for certain non-perturbative problems
within the non-interacting model. Since the inclusion of
interactions beyond the mean-field approach is impossible within
the supersymmetric method, the replica model the most reliable
tool for models with interactions.  The original fermionic replica
\SM\ \cite{EfLKh} has been generalized to include the interactions
by Finkelstein \cite{Fin}. It allowed him to reproduce earlier
perturbative results (reviewed, e.g., in Ref.~\cite{AA1}), to
account where necessary for the Landau Fermi-liquid constants, and
to derive the renormalization group equations describing (at least
at a qualitative level) a metal--insulator transition in
disordered interacting systems. This model has also been
successfully extended \cite{Fin:87} to allow for effects of
Coulomb interaction in superconducting systems (like lowering the
transition temperature $T_c$ by disorder -- in an apparent, but
well understood, deviation from the Anderson theorem
\cite{And:59}). The interest in this approach has been greatly
enhanced by the recent discovery \cite{Krav:95} of an apparent
metal-insulator transition in 2D disordered systems in zero
magnetic field. Although it is not at all clear whether the
observed effects are, indeed, due the transition
\cite{Sav+Pep2,Pep:1999}, and if they are -- whether such a
transition is, indeed, driven by interactions, a possibility of
having the transition in a 2D disordered interacting system
(schematically described by the TOE model \Ref{TOE}), in contrast
to a disordered noninteracting system, is intriguing by itself.
Such a possibility is, undoubtedly, a driving force in a
considerable revival of interest in Finkelstein's $\sigma$ model
(see, e.g., Refs.~\cite{AnKam:99,Lud:98,CdCKL:98,FLS:00,YL:01}).
At the moment, there is no clear evidence whether Keldysh
techniques  employed in  Refs.~\cite{AnKam:99,Lud:98} would give
any edge over the replica method used throughout this chapter.
Below I first describe how to include the Coulomb interaction (for
simplicity, in a singlet channel only, as the inclusion of a
triplet channel is technically almost identical) and the BSC
pairing interaction into the derivation of \NL\ given in section
\ref{replicas}. Then I will show (section \ref{SIT}) how to use
such a model for describing the superconductor-insulator
transition.

\subsection{Hubbard--Stratonovich decoupling}
Since the pair interaction does not conserve single-particle
energy, in contrast to the elastic scattering which was the only
mechanism included up to now, the effective interaction functional
should be dynamical. As usual, it is convenient to introduce
imaginary time $\tau$ implying everywhere the thermodynamic Gibbs
averaging (together with the averaging over quenched disorder
where applicable). Thus one considers thermodynamic Green's
functions related (in the Matsubara frequency representation) to
the retarded and advanced Green's functions of the previous
sections by the standard procedure of analytical continuation
\cite{AGD}. Then the effective functional corresponding to the
interaction term in \Eq{TOE} can be written as
\begin{equation}\label{EfFunc}
  \cS_{\text{int}}=\frac12\Int\dd x
  \dd
  x'\bar\psi_s(x)\bar\psi_{s'}(x')V_{xx'}\psi_{s'}(x')\psi_s(x)\,,
\end{equation}
where $x\equiv \b r, \tau$, and all the fermionic fields are
anti-periodic in imaginary time $\tau$ with period $1/T$,
$s=(\uparrow,\,\downarrow)$ is the spin index,
$V_{xx'}\equiv\d(\tau-\tau')V_c(\b r-\b r')$, and $V_c$ represents
the Coulomb interaction. After the replication, all the fields in
\Eq{EfFunc} naturally have the same replica index. Similarly, the
BCS functional describing the attraction in the Cooper channel can
be written as
\begin{equation}\label{BCSFunc}
\cS_{\text{sc}} =\lambda_0\int\!{\rm d}x\, \bar\psi_{\uparrow} (x)
\bar \psi_{\downarrow} (x) \psi_{\downarrow}(x)\psi_{\uparrow}(x),
\end{equation}
where $\lambda_0$ is the BCS coupling constant.

The Hubbard--Stratonovich decoupling of the functional
\Ref{EfFunc}-\Ref{BCSFunc} is similar to that for the disorder
functional \Ref{Sdis}. To allow for all slow modes in the
interaction functional\Ref{EfFunc} one should introduce three
matrix fields: the fields $\Phi$ and $\hat f$ to take account of a
small-angle scattering (a singlet channel) and large-angle
scattering, and yet another one that corresponds to the Coulomb
repulsion in the Cooper channel. The last one would lead to the
standard renormalization of the BCS attraction (see, e.g.,
Ref.~\cite{Nagaosa}) and is not considered here; further, it is
assumed that systems under considerations still have an effective
attraction in the Cooper channel after such a renormalization.
Under these conditions, the decoupling has the following form:
\begin{eqnarray}
\ee^{-\cS_{\text{int}}}&=&\Int\D\Phi\exp\left\{\!-\frac{1}{2}\!\Int\dd
x\dd x'\, \Phi(x) V^{-1}_{xx'}\Phi(x') +i\!\Int\dd
x\,\bar\psi_{s}(x)\Phi(x)\psi_{s}(x)\right\}
\nonumber\\+\;\quad&&\!\!\!\!\!\!\!\! \!\!\!\!\!\!\!\!\!\!\Int\D\!
\hat f\exp\left\{\!-\frac{1}{2}\!\Int\dd x\dd x'\,\tr\!\lr[]{\hat
f(x) V^{-1}_{xx'}\hat f(x')} +i\Int\dd
x\,\bar\psi_{s}(x)f_{ss'}(x)\psi_{s'}(x)\right\}\!,
\nonumber\\\label{Sfields}
\ee^{-\cS_{sc}}&=&\Int\D\Delta\exp\biggl\{-\frac{1}{\lambda}\int\dd
x \left|\Delta(x)\right|^2 \\\nonumber && +\,\,i\!\Int\dd
x\left[\Delta(x)\bar\psi_{\uparrow}(x)\bar\psi_{\downarrow}(x)
-\bar\Delta(x)\psi_{\downarrow}(x)\psi_{\uparrow}(x)\right]\biggr\}
\end{eqnarray}
Now one doubles the number of components of the fields $\psi $ and
$\bar \psi$  as in \Eq{CG} and performs the Gaussian integration
to obtain
\begin{equation}
\label{Str} \S=\S_{\text{fields}}+\frac{\pi\nu}{8\tau}\Tr \s^2 -
\frac{1}{2}\Tr\ln\left[ \x -i\left(\frac{1}{2\tau}\s
 + \Phi + \Delta +\hat\varepsilon\right)- \hat f\right].
\end{equation}
Here the operator $\hat \varepsilon$ equals $i \hat \tau_3
\partial _\tau$ in imaginary time representation and becomes the
diagonal matrix of fermionic Matsubara frequencies
($\varepsilon_n=\pi (2n+1)T$) in frequency representation;
${\mathrm Tr} $ refers to a summation over all the matrix indices
and to an integration over $\bbox r$ and $\tau$ (or summation over
Matsubara frequencies in the frequency representation);
$\S_{\text{fields}}$ includes the $\psi$-independent part of the
action \Ref{Sfields} quadratic in the fields $\Phi$, $\hat f$ and
$\hat \Delta$.  The triplet channel, described by the field $\hat
f$, is quite important: in particular, it can lead to the
delocalization in the presence of disorder \cite{Fin,CdCKL:98};
however, such effects will not be considered here and this term
will be ignored from now on  as it is not relevant for the
application of the model in section~\ref{SIT}. The field $\s$ has
the same structure as in \EQS{dc}{CG}, apart from explicitly
including the $2\times2$ spin sector and replacing the $2\times2$
retarded-advanced sector by the dependence on the imaginary time
$\tau$ (since it is diagonal in $\tau$, it becomes a matrix field
in the Matsubara frequencies). In the absence of the fields $\Phi$
and $\Delta$, the saddle-point solution for the action \Ref{Str}
is formally the same as in the zero-temperature techniques of
section~\ref{replicas} but the matrix $\L$ has a non-unit
structure in the Matsubara (instead of advanced-retarded) sector:
\begin{equation}\label{SP2}
 Q=U^\dagger\L U\,,\qquad \label{lambda}
\Lambda ={\text {diag\,\{sgn}}\,\hat\varepsilon\}\,.
\end{equation}
The field $\Phi$, corresponding to the singlet part of the Coulomb
interaction, is diagonal in all the sectors. The
``order-parameter'' field ${\hat \Delta}$ is   Hermitian and
self-charge-conjugate, diagonal in the replica indices and
coordinates $\mathbf r$ and $\tau$, and has the following
structure in the spin and time-reversal space:
\begin{equation}
{\hat \Delta}(x)= |\Delta(x)| e^{\frac{i}{2}\chi(x)\hat\tau_3}
\hat \tau^{sp}_2\otimes\hat\tau_2
e^{-\frac{i}{2}\chi(x)\hat\tau_3}\,, \label{Delta}
\end{equation}
where $|\Delta |$ and $\chi$ are the amplitude and the phase of
the pairing field $\Delta(\b r, \tau)$, $\hat\tau _i$ and
$\hat\tau^{\text{sp}}_i$ are Pauli matrices ($i=0,1,2,3$ with
$\hat\tau^{}_0=1$) that span the charge-conjugate and spin
sectors, respectively.

In the presence of the fields $\Phi$ and $\Delta$, the
saddle-point equation for the effective functional \Ref{Str} can
be formally written similarly to \Eq{SP} for the non-interacting
zero-temperature case:
\begin{equation}
\label{sp} -i\pi\nu\sigma({\mathbf r}) = \left\langle{\mathbf
r}\left|
\left[{-{\hat{\xi}}+ \frac{i}{2 \tau_{\text{el}}}\,\sigma+
i\left({
 \hat{\varepsilon}} + {\hat \Delta}+\Phi\right)}
\right]^{-1}\right|{\mathbf r}\right\rangle
\end{equation}
Now $\s=\L$  does no longer represent the saddle point solution
for $\eps\ne0$. Still, one can derive an effective functional by
expanding the above Tr ln within the manifold \Ref{SP2} in the
symmetry-breaking field $\eps+\Delta+\Phi$ and in gradients of
$Q$, as has been done in the original works by Finkelstein
\cite{Fin}. An alternative is to make first a similarity
transformation around $\L$ within the manifold \Ref{SP2} to find
the saddle point solution $\s_{\text{sp}}$ in the presence of the
fields. This can be formally done with the help of matrix $U_0$
that diagonalizes the Hermitian field $\hat\eps + \hat \Delta +
\Phi$:
\begin{equation}\label{diag}
\hat\eps + \hat \Delta + \Phi=U_0^\dagger \lambda U_0\,,\qquad
\s_{\text{sp}}=U_0^\dagger \L U_0\,.
\end{equation}
Here $U_0$ belongs to the same symmetry group that defines the
manifold \Ref{UQU} in the absence of the interaction fields. By
substituting \Eq{diag} into the saddle-point equation \Ref{sp},
one can easily verify that this is, indeed, a
spatially-homogeneous solution, provided that $\lambda \tel\E\gg1$
which will always be the case for a dirty superconductor
($\Delta\tel\E\gg1$) or a dirty metal (dimensionless conductance
$g\E\gg1$).

Now one can perform the expansion of Tr ln, \Eq{Str}, in gradients
of $Q$ and in the symmetry-breaking fields represented by the
eigenvalues $\lambda$ and the matrix $U_0$, \Eq{diag}. It is
convenient to  employ the following parameterization:
\begin{equation}
\label{s} \sigma= U_0^\dagger Q^{\,}U^{\,}_0\,, \qquad Q=
U^\dagger  \Lambda U\,,
\end{equation}
where $Q$ represents the saddle-point manifold in the metallic
phase and $\sigma$ is obtained from $Q$ by the same rotation
(\ref{diag}) as $\sigma_{\text{sp}}$ is obtained from the metallic
saddle point $\Lambda$. Therefore, $Q$ is defined, as in the
metallic phase, on the coset space $S(2N)/S(N)\otimes S(N)$ where,
depending on the symmetry, $S$ represents the unitary, orthogonal
or symplectic group as in the noninteracting case of
section~\ref{replicas}. The parameterization (\ref{s}) simplifies
considerably all the subsequent derivations and leads to a new
variant of the nonlinear $\sigma$ model which can be more
convenient for many applications than the original Finkelstein's
\NL. After substituting $\s=U_0^\dagger
U^{\dagger}_{\phantom{0}\,} \L^{\,}U^{\,}_0U^{\,}$, \Eq{s}, into
\Eq{Str}, one obtains the following representation for the
Tr$\,$ln term:
$$
\delta {\cal S} = -\tfrac 12 {\mathrm Tr}\ln \bigl\{\!\hat
G_0^{-1} +U^{\,}_ 0[\hat\xi,U^\dagger_0]-i(U\lambda U^\dagger)
\bigr\}\,,
$$
where one can also include an external magnetic field with the
vector potential $\b A$:
$$
\hat G_0\equiv \left( \hat \xi -\frac{
i}{2\tau_{\text{el}}}\Lambda \right)^{-1},\qquad
\x\equiv\frac1{2m}(\b p -e\b A)^2-\ef\,.
$$
The expansion to the lowest powers of gradients and $\lambda$ is
now straightforward and similar to that for the noninteracting
case. It results after some calculations in the following
effective action:
\begin{eqnarray}
{\cal S}= \frac{1}{2}\Tr\!\lr[]{ \Phi V^{-1}_{\b r\b r
'}\Phi}+\frac{1}{T\lambda_0}\sum_{\omega}\int\!\!{\mathrm d}{\bf
r}\,
|\Delta_{\omega}|^2 
+\frac{\pi\nu }{2} {\mathrm Tr}\!\left[\frac D4\left(\partial
Q\right)^2 - \lambda Q\right]\,. \label{new}
\end{eqnarray}
The long derivative in Eq.~(\ref{new}) is defined as
\begin{equation}
\partial Q
\equiv \nabla Q + \Bigl[{\bf A}_0 \!\!-\! {ie} {\bf A}\hat\tau_3,
\,Q \Bigr] \equiv \partial _0 Q + \left[{\bf A}_0 , Q \right] \,,
\label{LD}
\end{equation}
where the matrix ${\bf A}_0$ is given by
\begin{equation}
\label{A} {\bf A}_0 = U^{\,}_0\partial^{\,}_0 U^\dagger_0 \,,
\end{equation}
and $\partial_0\equiv\nabla-[ie\b A \hat\tau_3,\ldots] $ is the
long derivative (\ref{LD}) in the absence of the pairing field
$\Delta$. Both $U^{\,}_0$ and $\lambda$ should be found from the
diagonalization of $\epsilon + \Delta+\Phi$, Eq.~(\ref{diag}).
Although such a diagonalization cannot be done in general, it will
be straightforward in many important limiting cases. For
$\Delta=\Phi=0$, the field ${\bf A}_0$ vanishes,
$\partial\to\partial_0$, and $\lambda\to\varepsilon$, so that the
functional (\ref{new}) goes over to that of the standard nonlinear
$\sigma$ model for non-interacting electrons.

The action \Ref{new} is most general in the current context. It
allows one to develop a fully self-consistent approach to
superconductivity of dirty metals in the presence of Coulomb
interaction. However, any application requires a set of further
simplifications. As a simple illustration, I will show below how
to use the model in a dirty superconductor near the
metal-superconductor transition in the absence of Coulomb
interaction. Then, I will show in section \ref{SIT} how to use
this model to describe the insulator-superconductor transition as
a result of combined effects of disorder and interactions.

\subsection{Ginzburg-Landau Functional}

In the vicinity of the metal-superconductor transition one can
expand the action (\ref{new}) (in the absence of the Coulomb field
$\Phi$) in the pairing field $\Delta$. A further simplification is
possible in the weak disorder limit, $g\gg1$, when   the $Q$-field
can be integrated out to obtain an effective action for the
$\Delta$-field only. In the quadratic in $\Delta$ approximation,
the kernel of this action will give, with due account for the
disorder, an effective matrix propagator of the pairing field
which governs properties of a disordered superconducting sample
near the transition.

To integrate over the $Q$-field, one splits the action (\ref{new})
 into ${\cal S} \equiv {\cal S} _0 + {\cal S} _\Delta$ where
$\cS_0$ is the standard nonlinear $\sigma$ model functional as in
the metallic phase. Then one makes a cumulant expansion, i.e.\
first expands ${\rm e}^{-({\cal S} _ 0+ {\cal S} _\Delta)}$ in
powers of ${\cal S} _\Delta$, then performs the functional
averaging with ${\rm e}^{- {\cal S} _0}$ (denoted below by
$\langle\ldots\rangle_Q$) and finally re-exponentiates the
results. The expansion involves only the first and second order
cumulants since the higher order cumulants generate terms of
higher order in $\Delta$. Then the only contributions to the
action quadratic in $\Delta$ are
\begin{eqnarray}
\lefteqn{{\cal S}_{\rm eff}[\Delta] =\frac{1}{\lambda_0
T}\sum_{\omega}\int \!\!{\mathrm d}{\bf r}\, |\Delta_{\omega}|^2
-\frac{\pi\nu}{2}\Bigl\langle{\mathrm Tr}\, (\lambda
\!-\!\epsilon) Q\Bigr\rangle_{\!Q} }
\nonumber
\\[-3mm]
\label{av}
\\[-3mm]
\nonumber
&&-\biggl\langle\frac{\pi\nu D}{8}{\mathrm Tr}\,\left[{\bf
A}_0,Q\right]^2 +\frac{(\pi\nu D)^2}{8}\Bigl( {\mathrm Tr}\,
Q\partial _0 Q {\bf A}_0 \Bigr)^{\!2}\biggr\rangle_{\!\!Q}.
\end{eqnarray}
Expanding $\lambda$ and ${\bf A}_0$ to the lowest power in $\Delta
$ and performing the functional averaging  one finds \cite{YL:01}
the action quadratic in $\Delta$ as follows:
\begin{equation}
{\cal S}_{\rm eff}[\Delta] =\frac{\nu}{T}\sum_{\omega}\int\!{\rm
d} {\bf r}\, \Delta^*_{\omega}({\bf r})\bigl<{\bf
r}\bigl|\,\hat{\cal K}_{\omega} \,\bigr|{\bf r}'\bigr>
\Delta_{\omega}({\bf r}')\,, \label{K}
\end{equation}
with the operator $\hat{\cal K}_{\omega}$ given by
\begin{equation}
\label{wl}
\hat{\cal K}_{\omega}=\frac{1}{\lambda_0 \nu} -2\pi T\!\!\!\!
\sum_{\epsilon(\omega-\epsilon)<0} \left\{ \hat\Pi^{c}_{\omega}
+\frac{1}{\pi\nu} \frac{ \Pi^{d}_{|2\epsilon-\omega|}(0){\hat
{\cal C}}}{(2\epsilon-\omega)^2} \right\}.
\end{equation}
Here $\Pi^{c,d}_{|\omega|}({\bf r}, {\bf r}') =\bigl<{\bf
r}\bigl|\hat\Pi^{c,d}\bigr|{\bf r}'\bigr> $ are the cooperon and
diffuson propagators, with
\begin{equation}
\label{Pi}
\hat\Pi^{c}_{|\omega|} = \Bigl( {\hat{\cal C}} + |\omega|
\Bigr)^{-1}\,,
\end{equation}
where $ {\hat {\cal C}}\equiv -D\left(\nabla\E - {2ie}{\bf
A}\right)^2$ defines the cooperon modes; $\hat\Pi^{d}$ is obtained
from $\hat\Pi^{c}$ by putting the external vector potential $\bf
A=0$. In the last term in Eq.~(\ref{wl}), $\Pi^{d}_{|\omega|}
(0)\E\equiv \Pi^{d}_{|\omega|}({\bf r},{\bf r})$; this term may be
obtained by expanding in $g^{-1}$ the cooperon propagator with the
renormalized diffusion coefficient,
$$
{\hat{\cal C}} \to
\left[1-\frac{1}{\pi\nu}\Pi^{d}_{|\omega|}(0)\right]\,{\hat{\cal
C}},
$$
which is a weak localisation correction to free cooperon
propagator $\Pi^{c}_{|\omega|}({\bf r}, {\bf r}')$.

The summation over Matsubara frequencies in Eq.~(\ref{wl}) yields
\begin{equation}
\hat{\cal K}_{\omega}= \ln\frac{T}{T_0}+\psi\!\!\left(\!
\frac{1}{2}+\frac{|\omega|\!-\!{\hat{\cal C}} }{4\pi T}\! \right)
-\psi\!\!\left(\frac{1}{2}\right) - \frac{a_{\omega}{\hat{\cal
C}}}{4\pi T} \,, \label{K1}
\end{equation}
where $T_0\equiv T_{c0}(B\!=\!0)$ is the transition temperature of
the clean superconductor in the absence of a magnetic field and
$\psi$ is the digamma function. The weak localisation correction
is proportional to the coefficient $a_{\omega}$ given by
$$
\begin{array}{l}
\displaystyle a_{\omega}(T)=\frac{1}{\pi\nu
V}\sum_{\mathbf{q}}\frac{1}{Dq^2}\left\{
\psi'\left(\frac{1}{2}+\frac{|\omega|}{4\pi T}\right) \right. \\[6mm]
\displaystyle \left. -\frac{4\pi T}{Dq^2} \left[
\psi\left(\frac{1}{2}+\frac{|\omega|+Dq^2}{4\pi T}\right)
-\psi\left(\frac{1}{2}+\frac{|\omega|}{4\pi T}\right) \right]
\right\}.
\end{array}
$$
For $\omega=0$ the coefficient $a_0\equiv a_{\omega =0}(T)$ can be
simplified in the two limits:
\begin{equation}
 a_0=\left\{
\begin{array}{ll}
\displaystyle \quad\frac{\psi'({1/2})}{\pi\nu
L^d}\!\!\!\sum\limits_{L_{T}^{-1} <q<\ell^{-1}}\frac{1}{Dq^2}\,, &
L \gg L_T\,,\\[8mm]
\displaystyle -\frac{\psi''({1/2})}{8\pi^2\nu L^d T}\,\,, & L \ll
L_T \,,
\end{array}\right.
\end{equation}
where $L_T\equiv\sqrt{D/T}$ is the thermal smearing length.

The instability of the normal state (i.e. a transition into the
superconducting state) occurs when the lowest eigenvalue of the
operator ${\hat{\cal K}}_{\omega}$ becomes negative. The
eigenfunctions of this operator coincide with the eigenfunctions
of the cooperon operator ${\hat{\cal C}}$. The lowest eigenvalue
of ${\hat{\cal C}}$
 is known to be ${\cal C}_0= DB/\phi_0$, where
$\phi_0$ is the flux quanta. This ground state cooperon
eigenfunction corresponds to the lowest eigenvalue ${\cal K}_0$ of
the operator ${\hat{\cal K}}_{\omega}$. The condition ${\cal
K}_0=0$ implicitly defines the line $T_c(B)$ in the $(T,B)$-plane
where the transition occurs:
\begin{equation}
\ln\frac{T_c}{T_0} +\psi\left(\frac{1}{2}+\frac{{\cal C}_0}{4\pi
T_c}\right) -\psi\left(\frac{1}{2}\right)=\frac{a_0 {\cal
C}_0}{4\pi T_c}. \label{Tc(B)}
\end{equation}
The term in the r.h.s.\ of Eq.~(\ref{Tc(B)})
 describes a $1/g$-correction to the main result. This
 weak localisation is linear in the magnetic field $B$ and
vanishes as $B\to 0$ as expected (the Anderson theorem
\cite{And:59}). In a nonzero magnetic field the weak localisation
correction to $B_c$ is positive which has a very simple
explanation. The superconductivity is destroyed by the magnetic
field when $\Phi(\xi)\gsim\Phi_0$, where $\Phi(\xi)$ is the flux
over the area with linear size of the order of the coherence
length $\xi\sim\sqrt{D/T}$ and $\Phi_0$ is the flux quanta. The
weak localisation corrections reduce $D$ and thus  $\xi$.
Therefore, one needs a stronger field to destroy the quantum
coherence. The same reasoning explains the growth of $T_c$ in a
fixed magnetic field.

Note finally that it would be straightforward to include the
leading weak-localization corrections in all orders of $g^{-1}\ln$
by calculating the $Q$ - averages in \Eq{av} via the
renormalization group. This would lead to the renormalization of
$D$ in the cooperon propagator (\ref{Pi}), thus changing the shape
of the $T_c(B)$ curve. However, the value of $T_c(0)$ will again
remain unaffected, since the superconducting instability is
defined by the onset of the homogeneous zero mode in the operator
$\hat {\cal K}$, Eq.~(\ref{K1}), which does not depend on the
value of the diffusion coefficient in the cooperon propagator.

It is worth stressing that the Anderson theorem reflects certain
properties of a model rather than those of real superconductors.
If one allows for Coulomb interaction, then the critical
temperature of the superconducting transition is no longer
independent of disorder. Combined effects of the interaction and
disorder lead to corrections to the transition temperature
proportional (in a slightly simplified way) to $g^{-1}\ln
^3(T_c\tel)$ \cite{Ovchin:73,MEF:84,Fin:87}. For a relatively weak
disorder, the system still remains superconducting at $T\to0$
while for a sufficiently strong disorder the above corrections
would suppress the superconducting pairing at any temperature and
make a 2D system insulating at $T\to0$. Such a mechanism of
suppressing $\Delta$ by disorder gives a possible scenario for a
widely observed super\-conductor--insulator (SI) transition in
two-dimen\-sional structures [55--58] 
which is most adequately described within the \NL\
\cite{Fin:87,Fin} similar to those developed in this section.
However, a widely accepted scenario for such a transition is based
on very different, both in origin and in implementation, models
[59--63] where $|\Delta|$ remains finite and the superconductivity
is suppressed by the loss of the phase coherence due to the
fluctuations of the phase $\chi$ of the order parameter. I will
show in the following section that these ``phase-only'' models can
also be derived under certain parametrically controlled
assumptions from the \NL\ developed in this section.

\section{Superconductor--insulator transition in 2D systems:
phase and amplitude fluctuations of the order parameter
}\label{SIT}

A wide variety of experimental data indicate the existence of a
super\-conductor--insulator (SI) transition in two-dimen\-sional
films [55--58] 
(see, in
particular, \cite{GM:98} for earlier references).  The transition
can be tuned by either disorder (changing with the thickness of a
superconducting film) or magnetic field, thus being one of the
most intensely studied examples of quantum phase transitions
\cite{QPT:97}.

One of the most accepted ways to understand the problem of the SI
transition is based on the so-called Bose-Hubbard (or
``dirty-boson'') models \cite{FGG:90}. In this class of models
exists a duality between charge-$2e$ bosons (preformed Cooper
pairs) and vortices. The superconducting phase is due to the
bose-condensation of the charged bosons with localized vortices
while the insulating phase is due to the bose-condensation of the
vortices with the localized charged bosons. Another approach which
captures the basic physics of granular superconductors is based on
dissipative models \cite{CL:81} of resistively shunted charged
Josephson arrays
with the emphasis on the role of dissipation and Coulomb
interaction. In both group of models [59--65],
the transition is driven by fluctuations of the phase of
the order parameter, i.e.\ the superconductivity is destroyed in
spite of the existence of nonvanishing $|\Delta|$ locally.   This
approach seems to be rather different from that mentioned in the
previous section where the disorder and interaction destroy the
superconductivity by destroying $|\Delta|$ everywhere in a
homogeneously disordered system. An experimental distinction
between homogeneous and granular systems is not that
strict\cite{MK:99} as it seemed a few years ago, and recent
experimental observations \cite{MK:99,SIT:99a} strongly suggest
that the amplitude fluctuations in the vicinity of the SI
transition are no less important than the phase fluctuations.

The purpose of this section is to modify a general NL$\sigma$M
action for the description of granular systems with the BCS and
Coulomb interactions. Such a description takes account of
fluctuations of both amplitude and phase of the order parameter
$\Delta$, thus encompassing all the above described approaches.
Both the Bose-Hubbard model \cite{FGG:90} and the dissipative
models [59--61]
will be derived from this action via  certain controlled
simplifications made within the NL$\sigma$M. The latter model
 is more general and allows one to go beyond certain
limitations necessary in the derivation of the phase-only action.

The starting point is a coarse-grained version of the TOE model
Hamiltonian of \Eq{TOE} for a granular superconductor, where the
kinetic energy includes terms $t_{ij}a^\dagger_i
a^{\phantom{\dagger}}_j$ corresponding to tunnelling hops between
the grains. The derivation of the NL$\sigma$M from such
Hamiltonian follows the procedure described in the previous
section.  Thus a `half-baked' effective functional obtained by
integrating out the fermionic fields is similar to that in
\Eq{Str}:
\begin{eqnarray}
{\cal S}[\hat\sigma, \hat\Delta,\Phi]= \frac{\pi\nu}{8
\tau_{\text{el}}}\,{\mathrm Tr}\, \sigma ^2+
\frac1{4\lambda_{{\text{0}}}}{\mathrm Tr}\, |\hat\Delta|^2+\frac12
{\mathrm Tr}\,\Phi U^{-1} \Phi \nonumber
\\[-1mm]
\label{PDS}
\\[-1mm]
\nonumber -\frac12 {\mathrm Tr}\ln\left[ -{\hat{\xi}}-\hat t
+\frac{i}{2 \tau_{\text{el}}}\,\hat\sigma+ i\left( {\hat
\Delta}+\Phi +\hat  \epsilon \right )\right]\,.
\end{eqnarray}
The  difference here is that  $\hat \xi$ is the operator of the
intra-grain kinetic energy (counted from the chemical potential),
while $\hat t$ is the tunneling amplitude matrix (i.e.\ the
inter-grain kinetic energy). Note that in the granular case the
singlet field $\Phi$ is sufficient to decouple the Coulomb
interaction term so that the triplet field $f$ does not enter
\Eq{PDS}. All the bosonic fields have the same structure as before
but for the addition of the $m\times m$ grain sector. The symbol
Tr refers both to a summation over all these matrix indices and to
an integration over intra-grain position ${\bf r}$ and the
imaginary time $\tau$.

The principal simplification for granular systems is that all the
fields are spatially homogeneous inside each grain when the grains
are zero-dimensional, i.e.\ their
 sizes $L\alt  \xi, L_T $ ($\xi $ and $L_T$ are
 the superconducting and thermal coherence lengths)
which is equivalent to $|\Delta|, T\alt 1/\tau_{\text{erg}}$. For
a diffusive grain $\tau_{\text{erg}}=L^2/D\equiv
\hbar/E_{\text{\sc T}}$ while for a ballistic (chaotic) grain
$\tau_{\text{erg}}\sim L/\vf$. The tunneling matrix $\hat t
=\{t_{ij}\}$ depends only on grain indices, and the Coulomb
interaction thus reduces to the capacitance matrix, $U^{-1}\!\to
\!C_{ij}/e^2$.

To follow the procedure described in the previous section one
needs, first, to solve (formally) the saddle point conditions,
\EQS{diag}{s} separately for each grain.  As all the fields are
spatially homogeneous inside each grain, $U_0\equiv U_0(i)$
commutes with the operator $\hat \xi$. Then one only needs to
expand the Tr\,ln to the first nonvanishing orders in $t_{ij}$ and
$\lambda_i$, this expansion being justified when $|t|, |\Delta|,
T\ll 1/\tau_{\text{el}}\ll \ef$. Thus one arrives
 at the following effective action:
\begin{eqnarray}
{\cal S}[Q,\Delta,\Phi]= \int\limits_0^\beta{\rm d}\tau \biggl\{
\sum_i\frac{|\Delta_i|^2}{\nu\lambda_0\delta_i} +
\sum_{ij}\frac{C_{ij}}{2e^2}\,\Phi_i\Phi_j \biggr\}\nonumber
\\[-2mm]
\label{action}
\\[-2mm]
\nonumber - \sum_i\frac{\pi}{2\delta_i}{\mathrm Tr}\,\lambda_iQ_i
- \frac{g_{ij}^{\text{t}}}2 \sum_{ij}{\mathrm Tr}\,Q_iS_{ij} Q_j
S_{ji} \,,\quad
\end{eqnarray}
where $S_{ij}\equiv U^{\phantom{\dagger}}_0(i)U_0^\dagger(j)$, all
the fields depend on $\tau$, Tr refers  to all indices except
$i,\,j$ numerating grains, $\delta_i$ is mean level spacing in the
$i$-th grain, and the tunneling conductance is defined by
$g_{ij}^{\text{t}}\equiv 2\pi^2|t_{ij}|^2/\delta_i \delta_j$
(which is nonzero only for neighboring grains). Both $\lambda_i$
and $U_0(i)$ (and thus $S_{ij}$) should be found from the
diagonalization procedure in Eq.~(\ref{diag}), while $Q$ is given
by \Eq{s}.

The next step is to represent $U_0(i)$ as
\begin{equation}
\label{SV}
U_0(i)=V_i\, {\rm e}^{-\frac i2 \chi_i(\tau)\,\hat \tau_3}\,.
\end{equation}
This is similar to the gauge transformation suggested in
Refs.~\cite{KamGef:96} and used in Ref.~\cite{BEAH:00} to gauge
out the Coulomb field. However, one  cannot gauge out two
independent fields, $\Delta $ and $\Phi$. Substituting the
transformation (\ref{SV}) into the diagonalization condition
(\ref{diag}), one reduces it to
\begin{equation}
\label{diag2}
{\hat \epsilon}+\widetilde\Phi_i +{\hat \Delta}_i^{\!\text o}
=V_i^+\lambda_i V_i\,,
\end{equation} 
where ${\hat \Delta}_i^{\!\text o}$ is the field (\ref{Delta})
taken at $\chi =0$ and the field $\widetilde\Phi$ is given  in the
$\tau$ representation by $\widetilde\Phi_i\equiv {\Phi}_i- \tfrac
12 \partial _\tau \chi_i \,. $

Both $ \widetilde\Phi $ and $|\Delta |$ are massive fields whose
fluctuations are strongly suppressed. It is straightforward to
show that the fluctuations of $ \widetilde\Phi $ are of order
$\delta$ which is much smaller than both $T$ and $|\Delta|$.
 Therefore, in the mean field
approximation  in $ \widetilde\Phi $  this field can be neglected,
$ \widetilde\Phi=0$. This condition is nothing more than the
Josephson relation in imaginary time \cite{LO:83}. This locks the
fluctuations of the Coulomb field $ \Phi$ with the
phase-fluctuations of the pairing field $\hat \Delta$,
\begin{eqnarray}
\label{Phi}
\Phi=\tfrac12\partial _\tau \chi\,,
\end{eqnarray}
 thus reducing  the action (\ref{action})
 to one depending only on the fields $Q$
and $\Delta$.

The mean field approximation  in $|\Delta_i|$ is valid for
$|\Delta_i|\gg\delta$. It is equivalent to the standard
self-consistency equation which formally follows from the
variation of the action (\ref{action}) with respect to $|\Delta|$.
In this approximation one finds $|\Delta_i|$ to be independent of
$i$ and $\tau$. Thus the first term in Eq.\ (\ref{action}) becomes
a trivial constant, so that the action depends on  $Q$ and $\chi$
only:
\begin{eqnarray}
{\cal S}[Q,\chi]=\! \sum_{ij}\!\frac{C_{ij}}{8e^2}\!
\int\limits_0^\beta\!\!{\rm d}\tau\,
\partial _\tau \!\chi_i\,\partial _\tau \!\chi_j\!
-\! \sum_i\frac{\pi}{2\delta_i}{\mathrm Tr}\lambda_i Q_i \!-\!
\label{action2}
\frac{g_{ij}^{\text{t}}}2 \!\sum_{ij}{\mathrm
Tr}\,Q_iS_{ij}Q_jS_{ji}.\;
\end{eqnarray}
The field $\chi $ in this action obeys the  boundary condition
$\chi(\tau \E+\beta)= \chi(\tau)\,\text{mod}\,2\pi$. In
calculating the partition function, one should in general allow
for different topological sectors corresponding to different
winding numbers in $\chi$.

The ``phase-only'' action (\ref{action2}) includes neither
fluctuations of $|\Delta|$, nor fluctuations of $\Phi$ beyond the
Josephson relation, Eq.~(\ref{Phi}). Below this action will be
reduced to the AES action. Still, we stress that the action
\Ref{action2} is more general than the AES action. Thus, in the
absence of superconductivity, $\Delta \equiv 0$, it was shown
\cite{BEAH:00} that the former contains a correct screening of the
Coulomb interaction at low $T$, in contrast to the latter. This
may also be important in the case when $\Delta $ is much smaller
than the charging energy.

 To further simplify the action  (\ref{action2}) we note that
 the diagonalization conditions (\ref{diag2}), in the absence of
 the $|\Delta|$ fluctuations,
are the same for each grain and reduced to those solved in
Ref.~\cite{YL:01}:
\begin{eqnarray}
V_{\varepsilon\varepsilon'} =
\cos\frac{\theta_\varepsilon}2\,\delta_{\varepsilon,\varepsilon'}
+ \hat \tau^{sp}_2\!\otimes\!\hat\tau_2\,
\sin\frac{\theta_\varepsilon}{2}\,{\rm
sgn}\,\varepsilon\,\delta_{\varepsilon,-\varepsilon'} \nonumber
\\[-1mm]
\label{diagonal}
\\[-1mm]
\nonumber \lambda={\mathrm {diag}} \sqrt{\varepsilon^2   +
|\Delta|^2}\,{\mathrm {sgn}}\, \varepsilon \,,\quad
\cos\theta_\varepsilon\equiv
\frac{|\varepsilon|}{\sqrt{\varepsilon^2 + |\Delta|^2}} \,.
\end{eqnarray}
Then $S_{ij}$ in Eq.~(\ref{action2}) can be expressed in terms of
$V$ as
\begin{eqnarray}
\label{Sij}
S_{ij} \equiv V {\rm e}^{-\frac i2 \chi_{ij}\,\hat
\tau_3}V^\dagger\,, \qquad \chi_{ij}\equiv  \chi_{i} -
\chi_{j}\,.
\end{eqnarray}
Finally note that large-$|\varepsilon|$ contributions to the
action (\ref{SP}) are strongly suppressed, while for
$|\varepsilon|\ll |\Delta|$ one has $\lambda =|\Delta| \Lambda$
which suppresses fluctuations of $Q$ in each grain imposing
$Q\!=\!\Lambda$. Then, all matrices in the action (\ref{action2})
are diagonal in the replica indices so that these indices become
redundant.   It finally reduces the action \Ref{action2} to that
depending only on one {\it scalar} bosonic field $\chi_i(\tau)$,
the phase of the order parameter, which is a function of only the
imaginary time and the grain number. Indeed, the second term in
Eq.~(\ref{action2}) reduces to a trivial constant; evaluating the
tunneling term with the help of Eqs.~(\ref{diagonal}) and
(\ref{Sij}), one obtains
\begin{eqnarray}
\label{tunnel}
{\cal S}[\chi]&=& \sum_{ij}\biggl\{\frac{C_{ij}}{8e^2}
\int\limits_0^\beta\!\!{\rm d}\tau\,
\partial _\tau \!\chi_i\,\partial _\tau \!\chi_j
\\
&&-2{g_{ij}^{\text{t}}} \int\limits_0^\beta\!\!{\rm d}\tau\!\!
 \int\limits_0^\beta\!\!{\rm d}\tau '
g_{\text n}^2 (\tau \!-\!\tau ') \cos \chi^-_{ij}+
g_{\text a}^2 (\tau \!-\!\tau ') \cos \chi^+_{ij}
\biggr\}\nonumber ,
\end{eqnarray}
where $\chi_{ij}(\tau)\equiv\chi_{i}(\tau)-\chi_{j}(\tau)$,
$
 \chi_{ij}^\pm\equiv\tfrac12 \Bigl[
\chi_{ij}(\tau)\pm \chi_{ij}(\tau')\Bigr],
$ 
and the normal and anomalous Green's functions $g_{n,a}$
(integrated over all momenta)  are given by
\begin{eqnarray}
\label{GF}
g_n(\tau)\!=\!T\!\sum_{\varepsilon}\!\frac{\varepsilon\sin\varepsilon\tau}
{\sqrt{ \varepsilon^2 \!+\! |\Delta|^2}}\,, \;
g_a(\tau)\!=\!T\!\sum_{\varepsilon}\!\frac{|\Delta|\cos\varepsilon\tau}
{\sqrt{ \varepsilon^2 \!+\! |\Delta|^2}}\,.
\end{eqnarray}
The action \Ref{tunnel} is exactly the AES action introduced in
Refs.~\cite{AES:82,LO:83,AES:84}. Further simplifications are
possible in two limiting cases.

First,  in the normal case ($\Delta\!=\!0$ one has in
Eq.~(\ref{GF})
\begin{eqnarray}
\label{GF2}
g_a=0\,,\qquad g_n^2(\tau) =\frac{T^2}{\sin^2\pi T\tau}\,.
\end{eqnarray}
Then the field $\chi$ should be substituted, according to
Eq.~(\ref{Phi}), by $2\int^\tau\!{\rm d}\tau' \Phi(\tau')$. This
corresponds to using the action (\ref{tunnel}), (\ref{GF2}) for a
set of normal tunnel junctions \cite{SZ}, as has been recently
shown in Ref.~\cite{BEAH:00}; the functional (\ref{action2}) in
the limit $\Delta=0$ is equivalent to that of Ref.~\cite{BEAH:00}.
Including disorder-induced fluctuations (i.e.\ going beyond the
$Q=\Lambda$ approximation) allows one to obtain \cite{BEAH:00} a
correct low-$T$ limit for the phase correlation function missing
in the action (\ref{tunnel}).

The action (\ref{action2}) is more general than that considered in
Ref.~\cite{BEAH:00}: although under the mode locking condition
(\ref{Phi})  it depends only on the fields $\chi$ and $Q$, the
matrix $S_{ij}$, Eqs.~(\ref{diagonal}) and (\ref{Sij}), reduces to
a simple $U(1)$ gauge transformation as in in  Ref.~\cite{BEAH:00}
only in the limit $\Delta=0$.

The second limiting case, $T\ll |\Delta|$, is just the limit
relevant in the context of the SI transition in granular
superconductors. For $T=0$, the summation in Eq.~(\ref{GF}) can be
substituted by integration which yields
\begin{eqnarray*}
g_n(\tau)=\frac{|\Delta|}{\pi}\,K_1(|\Delta|\tau), \quad
g_a(\tau)=\frac{|\Delta|}{\pi}\,K_0(|\Delta|\tau)
\end{eqnarray*}
This is also a good approximation for a low-$T$ case; substituting
this into Eq.\ (\ref{tunnel}) gives the action for the dissipative
model\cite{LO:83,AES:84}. Note that for
$|\tau-\tau'|\ll|\Delta|^{-1}$, the main contribution in the
tunneling action (\ref{tunnel}) is given by the normal term with
the corresponding kernel proportional to $|\tau-\tau'|^{-2}$. The
Fourier transform of this would give a term of the
Caldeira-Leggett type\cite{CL:81} proportional to $|\omega|$.

The tunneling action (\ref{tunnel}) is non-local in $\tau$. As has
been noted in Ref.~\cite{LO:83} for the case of one Josephson
junction, for sufficiently large capacitance the phase $\chi_{ij}$
changes slowly in comparison with $|\Delta|^{-1}$, and in the
adiabatic approximation $\chi( \tau')$ is changed by $\chi(\tau)
+(\tau'\!-\!\tau)\partial_ \tau\chi(\tau)$. Making such an
expansion, one obtains from Eq.~(\ref{tunnel}) the following local
action:
\begin{eqnarray}
\label{DB}
{\cal S}[\chi]= \int\limits_0^\beta\!\!{\rm d}\tau\,\biggl\{
\sum_{i j} \tfrac12u^{-1}_{ij} \dot \chi_i \dot \chi_j -|\Delta|
{g_{ij}^{\text{t}}} \cos \chi_{ij} \biggr\} ,
\end{eqnarray}
where $\dot \chi_i\equiv \partial _\tau \!\chi_i$ and
\begin{eqnarray*}
\frac1{u_{ii}}&\equiv& \frac{C_{ii}}{4e^2}
+\sum_j\frac{{g_{ij}^{\text{t}}}}
{|\Delta|}\frac{3+\cos \chi_{ij}}8\,,\\
u^{-1}_{ij} &\equiv& \frac{C_{ij}}{4e^2}
-\frac{{g_{ij}^{\text{t}}}} {|\Delta|}\frac{3+\cos \chi_{ij}}8\,.
\end{eqnarray*}
If all the self-capacitances are equal to $C$ with $E_c\propto
e^2/C$ being the charging energy, and all
${g_{ij}^{\text{t}}}={g^{\text{t}}}$, then $u_{ii}\equiv U$ has
the meaning of the renormalized charging energy. Ignoring
 a weak dependence of $u$ on $\cos \chi_{ij}$
in  the above relations, one obtains the renormalized charging
energy:
\begin{eqnarray}
\label{U}
U=\frac {E_c}{1+\# E_c {g^{\text{t}}}/|\Delta|  }\,.
\end{eqnarray}
Here the coefficient $\#$ depends on the number of next neighbors
for each grain, etc. A similar renormalization takes place for the
next-neighbor off-diagonal energy $u_{ij}$. Now one can see that
on the face of it the adiabatic approximation employed to obtain
Eq.~(\ref{DB}) is valid for $U\ll |\Delta| $. However, in the
region $g^{\text{t}}\gg |\Delta|/E_c$, where the charging energy
(\ref{U}) is strongly renormalized, the instanton-like solutions
\cite{Korshunov} may be important. This may further reduce the
region of applicability for the local in $\tau$ action (\ref{DB}).

Finally, by introducing the operator $\hat n$ canonically
conjugate to the phase $\chi$, one finds the Hamiltonian that
corresponds to the action (\ref{DB}):
\begin{eqnarray}
\label{HDB}
\hat H= \sum_{i j} \tfrac12u_{ij} \hat n _i \hat n_j -|\Delta|
{g_{ij}^{\text{t}}} \cos\left( \chi_{i}-\chi_j\right)\, .
\end{eqnarray}
This is just the Hamiltonian of the Bose-Hubbard model
\cite{FGG:90} which was first microscopically derived by Efetov
\cite{Ef:80} in the context of granular superconductors.

To conclude, in this section the effective \SM-type action
(\ref{action}) has been derived for a granular system with
zero-dimensional grains in the presence of the Coulomb and pairing
interactions. This is the most general (in the present context)
action which takes into account fluctuations of both amplitude and
phase of the order parameter $\Delta$. Neglecting fluctuations of
$|\Delta|$ and fluctuations of the Coulomb field $\Phi$ beyond the
Josephson relation (\ref{Phi}) reduces this action to the
"phase-only" action (\ref{action2}) which still contains
intra-granular disorder important for the correct screening for
$|\Delta|$ small compared to the charging energy. Neglecting this
disorder further reduces the action (\ref{action2}) to that of the
AES model, Eq.~(\ref{tunnel}). When the renormalized charging
energy, Eq.~(\ref{U}), is much smaller than $|\Delta|$, the action
(\ref{tunnel}) finally goes over to that of the Bose-Hubbard
model, Eq.~(\ref{HDB}), which is widely used for the description
of the superconductor-insulator transition\cite{FGG:90}. However,
the above estimations show that this reduction is parametrically
justified only for  the region $E_c\ll |\Delta|$ where the
transition happens at $g^{\text{t}}\ll E_c/ |\Delta|\ll1$ which
corresponds to a strongly granular system. Note finally that the
most general (in the present context)  action (\ref{action})
describes both amplitude and phase fluctuations of the order
parameter, being still considerably different from the NL$\sigma$M
action for homogeneous systems.

{\bf Acknowledgments.} This work has been supported by the
Leverhulme  Trust under the contract F/94/BY and by the EPSRC
grant GR/R33311.


\end{document}